\documentclass[journal]{IEEEtran}
\usepackage{acro} 
\DeclareAcronym{SPC}{short = SPC, long = short-packet communications, tag = abbrev}
\DeclareAcronym{MR-BMOCZ}{short = MR-BMOCZ, long = multi-radius binary modulation on conjugate-reciprocal zeros, tag = abbrev}
\DeclareAcronym{TO}{short = TO, long = timing offset, tag = abbrev}
\DeclareAcronym{OFDM}{short = OFDM, long  = orthogonal frequency division multiplexing, tag   = abbrev}
\DeclareAcronym{OOK}{short = OOK, long  = on-off keying, tag   = abbrev}
\DeclareAcronym{IoT}{short = IoT, long = Internet of things, tag = abbrev}
\DeclareAcronym{CFO}{short = CFO, long = carrier frequency offset, tag = abbrev}
\DeclareAcronym{mMTC}{short = mMTC, long = massive Machine-Type Communications, tag = abbrev}
\DeclareAcronym{CIR}{short = CIR, long = channel impulse response, tag = abbrev}
\DeclareAcronym{MOCZ}{short = MOCZ, long = modulation on conjugate-reciprocal zeros, tag = abbrev}
\DeclareAcronym{BMOCZ}{short = BMOCZ, long = binary modulation on conjugate-reciprocal zeros, tag = abbrev}
\DeclareAcronym{S-BMOCZ}{short = S-BMOCZ, long = Smooshed BMOCZ, tag = abbrev}
\DeclareAcronym{PAPR}{short = PAPR, long = peak-to-average power ratio, tag = abbrev}
\DeclareAcronym{J-BMOCZ}{short = J-BMOCZ, long = Jutted-BMOCZ, tag = abbrev}
\DeclareAcronym{DiZeT}{short = DiZeT, long = direct zero test, tag = abbrev}
\DeclareAcronym{LTI}{short = LTI, long = linear-time invariant, tag = abbrev}
\DeclareAcronym{ERLLC}{short = ERLLC, long = extremely reliable and low-latency communication, tag = abbrev}
\DeclareAcronym{URLLC}{short = URLLC, long = ultra-reliable low latency communication, tag = abbrev}
\DeclareAcronym{BER}{short = BER, long = bit error rate, tag = abbrev}
\DeclareAcronym{BLER}{short = BLER, long = block error rate, tag = abbrev}
\DeclareAcronym{AWGN}{short = AWGN, long = additive white Gaussian noise, tag = abbrev}
\DeclareAcronym{RMSE}{short = RMSE, long = root mean square error, tag = abbrev}
\DeclareAcronym{CCDF}{short = CCDF, long = complementary cumulative distribution function, tag = abbrev}
\DeclareAcronym{SNR}{short = SNR, long = signal-to-noise ratio, tag = abbrev}
\DeclareAcronym{CP}{short = CP, long = cyclic prefix, tag = abbrev}
\DeclareAcronym{IM}{short = IM, long = index modulation, tag = abbrev}
\DeclareAcronym{AACF}{short = AACF, long = aperiodic autocorrelation function, tag = abbrev}
\DeclareAcronym{NMSE}{short = NMSE, long = normalized mean square error, tag = abbrev}

\usepackage{amsmath,amssymb,amsfonts}
\usepackage{mathtools}
\usepackage{array}    
\usepackage{tabularray}
\usepackage{graphicx}
\usepackage{textcomp}
\usepackage{xcolor}
\usepackage{listings}
\usepackage[noend]{algpseudocode}
\usepackage{pifont} 
\usepackage{subcaption}
\usepackage{caption}
\definecolor{GreenForest}{rgb}{0.09, 0.45, 0.27}
\usepackage{soul}
\usepackage{balance}
\usepackage{cite}
\usepackage{float}
\usepackage{algorithm}
\usepackage{algpseudocode}
\usepackage[colorlinks=true, linkcolor=blue, citecolor=blue, urlcolor=blue]{hyperref}
\def\BibTeX{{\rm B\kern-.05em{\sc i\kern-.025em b}\kern-.08em
T\kern-.1667em\lower.7ex\hbox{E}\kern-.125emX}}

\begin{document}
\bstctlcite{IEEEexample:BSTcontrol}
\title{Synchronization-Oriented Multi-Radius BMOCZ for Non-Coherent Short-Packet Communications}
\author{Hamoud Abdelbasset Hamidouche, Riadh Ouzane, Salah Eddine Zegrar,~\IEEEmembership{Member,~IEEE},  and H\"{u}seyin Arslan,~\IEEEmembership{Fellow,~IEEE}

\thanks{Hamoud Abdelbasset Hamidouche, Riadh Ouzane and Hüseyin Arslan are with the Department of Electrical and Electronics Engineering, Istanbul Medipol University, Istanbul, 34810, Turkey (e-mail: hamidouche.hamoud@std.medipol.edu.tr; ouzane.riadh@std.medipol.edu.tr).
Salah Eddine Zegrar is an independent
researcher (e-mail: salah.zegrar@shaynets.com.tr).}}

\maketitle

\begin{abstract}
Modulation on conjugate-reciprocal zeros (MOCZ) is a promising candidate for blind short-packet communication (SPC) in future wireless systems; however, its performance is sensitive to timing and frequency synchronization errors. In this paper, we propose a novel scheme named multi-radius binary MOCZ (MR-BMOCZ). This method introduces a new root placement strategy that enhances diversity in the autocorrelation domain. The resulting structure enhances the autocorrelation properties of the transmitted waveform, leading to improved timing offset (TO) estimation. Moreover, the multi-radius construction distributes the signal energy over a larger set of radii, hence reducing the peak-to-average power ratio (PAPR). The resulting waveform also exhibits a distinctive spectral structure that is exploited for carrier frequency offset (CFO) estimation and correction, yielding lower normalized mean square error (NMSE) compared with BMOCZ-based schemes. Simulation results over additive white Gaussian noise (AWGN) and fading channels demonstrate that MR-BMOCZ achieves superior synchronization performance while maintaining comparable bit error rate (BER) and block error rate (BLER) performance. Consequently, the proposed scheme offers a synchronization-oriented extension of BMOCZ for robust non-coherent SPC.
\end{abstract}

\begin{IEEEkeywords}
Short-packet communication, BMOCZ, timing offset, PAPR, carrier frequency offset.
\end{IEEEkeywords}

\section{Introduction}

\IEEEPARstart{T}{he} transition toward sixth-generation (6G) wireless networks is expected to extend the scope of 5G \ac{URLLC} to meet the growing demands of \ac{ERLLC}  \cite{bhat20216g,ji2018ultra}. This evolution is driven by a wide range of \ac{IoT} applications, including massive and mission-critical machine-type communications such as industrial automation, autonomous control, vehicular networks, and medical telemetry \cite{simsek20165g,siddiqui2023spectrally}. Such applications generally involve the transmission of short data packets under extremely tight latency and reliability constraints \cite{durisi2016toward}. In the coherent short-packet regime, i.e., \ac{OFDM}-based systems, the use of short symbol durations typically necessitates an increased subcarrier spacing \cite{godard2003self}, which in turn increases sensitivity to channel frequency selectivity. To preserve orthogonality under such conditions, a longer \ac{CP} is often required, resulting in spectral inefficiency. Furthermore, pilot-assisted channel estimation and tracking can consume a non-negligible fraction of the packet duration, thereby degrading latency, energy efficiency, and achievable throughput \cite{jia2025coherent,jia2025optimum}. Therefore, non-coherent communication has become an alternative for short-packet and highly dynamic wireless systems \cite{godard2003self}. By avoiding pilot overhead and channel tracking, non-coherent schemes reduce receiver complexity and are suited to scenarios where the channel coherence time is insufficient to justify training-based approaches \cite{9281072,bockelmann2016massive,witrisal2009noncoherent}. However, the removal of channel knowledge shifts the system bottleneck toward blind synchronization and detection under uncertainty, challenges that are critical in short-packet transmissions \cite{wang2023covert,kuscu2019transmitter}. Several works have investigated non-coherent transmission strategies tailored to the short-packet regime, such as non-coherent \ac{IM}-\ac{OFDM}, which was introduced in \cite{choi2017noncoherent}, and \ac{OFDM}-\ac{OOK} in \cite{sahin2018sequence}, which has been considered within the context of ambient \ac{IoT} and low power wake-up signal systems due to its low-complexity receiver architecture and energy-efficient operation \cite{3gppTR38769, zegrar2026ofdm}. While these approaches eliminate the need for explicit channel estimation, they remain affected by channel distortions in the frequency domain and cannot fully exploit multipath diversity\cite{walk2017short, dong2025time}. This residual channel dependence motivates the exploration of alternative signaling domains in which the impact of the channel is minimized. 

Recently, a novel modulation paradigm, denoted as \ac{MOCZ}, has been proposed to address these challenges in a systematic manner \cite{walk2019mocz}. In \ac{MOCZ}, information is encoded in the complex zeros of the $z$-transform of the discrete-time transmit signal, yielding a polynomial-based representation \cite{eren2025generalized}. The primary advantage of this framework lies in its multipath modeling. Unlike conventional multicarrier systems, where the channel effect is represented through circular convolution, \ac{MOCZ} models transmission in the $z$-domain such that linear convolution in the time-domain corresponds to polynomial multiplication \cite{perre2025learning, eren2024dynamic}. Under this representation, the channel introduces its own zeros into the received polynomial while the information-bearing zeros remain structurally identifiable. As a result, the received signal contains both channel-induced and data-carrying zeros, allowing recovery directly from the received zero structure without explicit channel estimation \cite{sun2023noncoherent}. This decoupling between channel and information-bearing zeros provides robustness against unknown multipath propagation and makes \ac{MOCZ} particularly attractive for blind non-coherent \ac{SPC} scenarios\cite{huffman2003generation}. Among the \ac{MOCZ} family, \ac{BMOCZ} has received particular attention due to its binary mapping strategy based on conjugate-reciprocal zero pairs. In this scheme, the placement radius of the zeros in the complex plane plays a central role, as it determines inter-zero distances, codeword separation, and ultimately detection reliability \cite{huggins2024optimal}. Prior work has identified an optimal single-radius choice that maximizes the minimum zero separation and proposed the corresponding \ac{DiZeT} detection framework for blind non-coherent reception in both time and frequency domains \cite{walk2019mocz}. The \ac{BMOCZ} scheme produces transmit sequences that belong to the class of Huffman sequences, which are known for near-impulsive autocorrelation properties \cite{svalbe2020families}. These properties are beneficial in noisy environments and enable efficient blind synchronization, channel-length estimation, and robust sensing capabilities \cite{dehkordi2023integrated}. Recent advances have extended \ac{MOCZ} to a range of emerging scenarios, including spectrally efficient frameworks such as faster-than-Nyquist signaling, \ac{PAPR}-aware codebook designs, and multiuser transmission schemes \cite{siddiqui2023spectrally,sasidharan2024alternative,eren2025non}. Despite these developments, existing \ac{MOCZ} designs remain largely restricted to single-radius constructions. Although the use of multiple radii has appeared implicitly in early \ac{MOCZ} formulations where random radii within an annulus are employed to obscure the zero codebook \cite{walk2018physical}, the radial variation in these schemes is neither structured nor optimized for waveform design. Instead, it serves as a physical-layer security-oriented randomization mechanism, without controlled radius allocation or analysis of the resulting synchronization and signal properties. The single-radius zero structure provides limited geometric diversity for distinguishing \ac{TO} and \ac{CFO} effects, while random radial placement offers no deterministic structure from which synchronization-related features can be systematically extracted or optimized. Moreover, the impact of controlled radial placement on waveform properties such as \ac{PAPR}, autocorrelation, \ac{TO} and \ac{CFO} identifiability remains largely unexplored. To the best of our knowledge, a deterministic multi-radius zero architecture that exploits radial placement as a systematic design degree of freedom for communication and synchronization has not been investigated.

Motivated by this gap, this paper presents \ac{MR-BMOCZ} as a suitable candidate for \ac{SPC}, considering practical challenges such as \ac{TO}, \ac{CFO}, and \ac{PAPR}. The proposed scheme distributes information across multiple zero radii and represents the transmitted waveform as the product of several lower-degree \ac{BMOCZ} polynomials, introducing additional diversity in the autocorrelation domain and thereby addressing key limitations of conventional schemes. 

Accordingly, the main contributions of this work are summarized as follows:

\begin{itemize}

\item We propose a novel multi-radius extension of \ac{BMOCZ}, named \ac{MR-BMOCZ}, in which information-bearing zeros are distributed across multiple radius pairs in the $z$-plane, introducing an additional design degree of freedom while preserving the \ac{BMOCZ} structure.

\item We establish the autocorrelation properties of \ac{MR-BMOCZ} and show that the convolution of multiple Huffman autocorrelations generates additional deterministic peaks, yielding a richer synchronization signature than conventional \ac{BMOCZ}. The resulting structure is characterized in both the time and $z$-domains.

\item By exploiting this autocorrelation diversity, we develop \ac{TO} estimation techniques based on a multi-peak timing metric and a generalized Huffman bracket. The resulting synchronization framework is evaluated through \ac{BER} and \ac{BLER} simulations under timing synchronization errors

\item We show that the multi-radius construction produces a distinctive spectral structure for \ac{CFO} estimation. The condition for unambiguous \ac{CFO} estimation is established, and a generalized multi-radius \ac{DiZeT}-based decoder is developed and evaluated through \ac{NMSE} analysis.

\item We analyze the impact of the multi-radius construction on waveform characteristics and detection performance. Distributing signal generation across lower-degree sub-polynomials reduces time-domain energy concentration and improves \ac{PAPR}, while root-perturbation analysis examines the sensitivity of the zeros to noise.

\end{itemize}

The remainder of this paper is organized as follows: Section \ref{preliminaries} reviews \ac{BMOCZ} and Huffman sequences. Section \ref{sysmodelsection} presents the proposed \ac{MR-BMOCZ} and receiver-side processing. Section \ref{sec:sync} investigates autocorrelation diversity and the corresponding \ac{TO} and \ac{CFO} estimation methods. Section \ref{Performance} provides the theoretical analysis, and Section \ref{simulationsection} presents simulation results. Finally, Section \ref{conc} concludes the paper\footnote{\emph{Notation:} Bold uppercase $\mathbf{A}$, bold lowercase $\mathbf{a}$, and unbolded letters $A,a$ denote matrices, column vectors, and scalars, respectively. The upright calligraphic letter (e.g., $\mathcal{C}$) denotes a set or codebook. Basic operators include the Kronecker delta $\delta[\cdot]$, convolution $*$, direct-sum operation $\oplus$, and complex conjugation $\overline{x}$. For vectors $\mathbf{x}\in\mathbb{C}^N$, $|\mathbf{x}|_p$ denotes the $\ell_p$-norm. The notation $\mathbf{x}^{-}$ denotes the time-reversed conjugate vector.}.

\section{PRELIMINARIES ON HUFFMAN-BMOCZ}
\label{preliminaries}
\ac{BMOCZ} relies on the concept of modulating the information onto the zeros of the baseband signal's z-transform. Considering a binary message of length $K$ bits $\mathbf{m}=[m_0,m_1,\ldots,m_{K-1}], \quad m_k\in\{0,1\}$, the $k$-th message bit is mapped to the zero of a polynomial as 
\begin{equation}
\alpha_k =\begin{cases}
\alpha_k^{(1)} = R e^{j\phi_k},     & m_k = 1\\
\alpha_k^{(0)} = R^{-1} e^{j\phi_k}, & m_k = 0
\end{cases}
\quad, k=0,1,\cdots,K-1,
\label{map}
\end{equation}
where $R$ and $R^{-1}$ denote the radii of the outer and inner circles, respectively. Following~\cite{walk2018physical}, the radius is chosen as
$R=\sqrt{1+\sin\left(\frac{\pi}{K}\right)}$ to improve noise robustness, while the phases are uniformly distributed as $\phi_k=\frac{2\pi k}{K}$.
The mapping from the designed zero constellation \eqref{map} to the time-domain sequence is established through the fundamental theorem of algebra~\cite{walk2019mocz}. Specifically, any finite-length discrete-time sequence can be represented by a polynomial whose coefficients correspond to the signal samples, and this polynomial can be factorized into the product of its complex zeros, given by
\begin{equation}
\label{X(z)_K}
X(z) = x_K \prod_{k=0}^{K-1} (z - \alpha_k) = \sum_{k=0}^{K} x_k z^k,
\end{equation}
where $x_K\neq0$ denotes the leading coefficient and is chosen to satisfy the energy normalization constraint $\|\mathbf{x}\|_2^2=K+1$. The polynomial $X(z)$ is the z-transform of the sequence $\mathbf{x}=[x_0,x_1,\ldots,x_K] \in \mathbb{C}^{K+1}$. This representation establishes a one-to-one mapping between the signal coefficients and the zero constellation, thereby forming the foundation of \ac{BMOCZ} signaling, where information is encoded in the locations of the zeros rather than in the signal amplitudes or phases.

Considering an $L$-tap frequency-selective channel, the received signal can be expressed as
\begin{equation}
    y[n] = \sum_{m=0}^{L-1}h_m x[n-\ell_m]+w[n],
    \label{eq:TD}
\end{equation}
where $h_m \sim \mathcal{CN}(0,\rho^{m})$ denotes the complex gain of the $m$-th channel tap, and $\rho \leq 1$ is the exponential power decay factor, such that the average power of the $m$-th tap is given by
$\mathbb{E}\{|h_m|^2\}=\rho^{m}$.
Furthermore, $\ell_m$ denotes the delay of the $m$-th propagation path, and $w(n)\sim\mathcal{CN}(0,N_0)$ is the \ac{AWGN}. Since linear convolution in the time-domain corresponds to polynomial multiplication in the z-domain \cite{oppenheim1997signals},
the equivalent representation of \eqref{eq:TD} can be expressed as
\begin{equation}
\begin{aligned}
Y(z) & = X(z)H(z) + W(z) \\
 & = x_K h_{L-2} 
\prod_{k=0}^{K-1} (z - \alpha_k)
\prod_{l=0}^{L-1} (z - \beta_l) \\
&\quad+ \underbrace{w_{N-1} 
\prod_{n=0}^{N-2} (z - \gamma_n)}_{\text{noise perturbation}}.
\end{aligned}
\label{eq:ZD}
\end{equation}
It can be observed from~\eqref{eq:ZD} that, due to the $L$-tap channel, the received signal in the $z$-domain contains additional roots while preserving the original information-bearing zeros of $X(z)$. The total number of roots is $N=K+L-1$, among which $K$ correspond to data zeros, perturbed by the additive noise.

\subsubsection{Key Property}
A key property of Huffman-BMOCZ is its almost-perfect \ac{AACF}, which exhibits an impulse-like structure and can be exploited for \ac{TO} estimation at the receiver~\cite{csahin2024over}.

For any BMOCZ sequence $\mathbf{x}\in\mathbb{C}^{K+1}$, the normalized \ac{AACF} is given by
\begin{equation}
\begin{aligned}
\mathbf{a}_\mathrm{B}(K,\eta)
&= \mathbf{x} \ast \overline{\mathbf{x}^{-}}
= \bigl(-\eta,\; \mathbf{0}_K,\; 1,\; \mathbf{0}_K,\; -\eta \bigr),
\label{impulsive}
\end{aligned}
\end{equation}
where $\eta = \bigl(R^K + R^{-K}\bigr)^{-1}$.
It follows that the \ac{AACF} contains only three nonzero coefficients: a dominant central peak of unit magnitude and two symmetric side peaks of magnitude $\eta$. This sparse autocorrelation structure in Huffman-BMOCZ enables robust synchronization and \ac{TO} estimation.

\section{PROPOSED MULTI-RADIUS BMOCZ MODULATION}
\label{sysmodelsection}
This section presents the proposed multi-radius BMOCZ modulation scheme together with the associated receiver-side signal processing procedure.

\subsection{Multi-radius BMOCZ Modulation}

\begin{figure*}[t!]
\centering
\resizebox{2\columnwidth}{!}{
\includegraphics{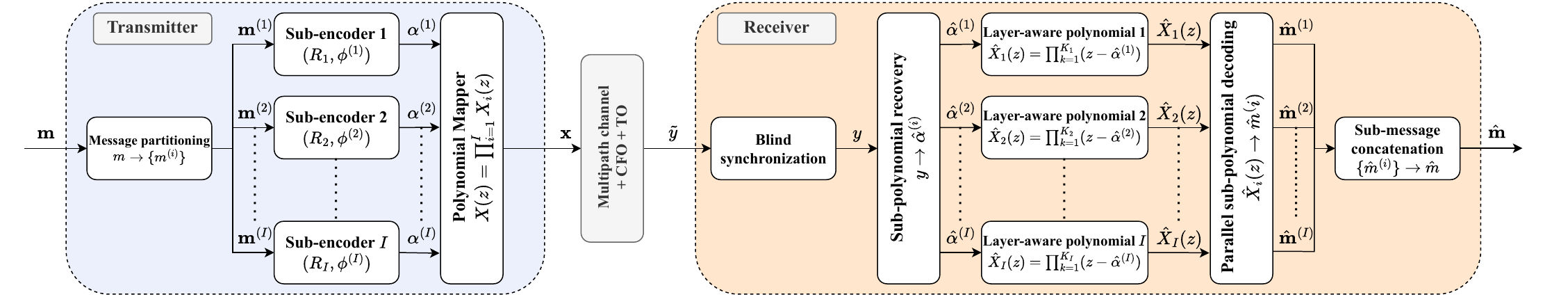}}
\caption{Block diagram representation of the proposed MR-BMOCZ}
\label{blockdiagram}
\end{figure*}

The high-level structure of the proposed scheme is as follows. The $K$-bit information sequence is
partitioned into $I>1$ submessages
\begin{equation}
\mathbf{m} =
\big[\mathbf{m}^{(1)},\, \mathbf{m}^{(2)}, \dots, \mathbf{m}^{(I)}\big],
\end{equation}
where the submessage lengths may differ across realizations and satisfy
\begin{equation}
\left\{
\begin{aligned}
\big|\mathbf{m}^{(i)}\big| &= K_i,\quad i=1,\ldots,I,\\
\sum_{i=1}^{I} K_i &= K.
\end{aligned}
\right.
\end{equation}
The number of bits assigned to each submessage may be equal or different, depending on the realization, as long as it holds that the number of bits across all the submessages  eventually adds up to $K$. 
In particular, the $k$-th zero associated with the $i$-th submessage is placed as
\begin{equation}
\label{eq:set}
    \alpha_{k}^{(i)} = R_i^{\,2m^{(i)}_k-1} e^{j\theta_k^{(i)}}, \qquad m^{(i)}_k\in\{0,1\},
\end{equation}
with 
\begin{equation} \label{eq8}
\begin{aligned}
    & R_{i} = \sqrt{1+\sin\left(\frac{\pi}{K_i}\right)}, \\
    & \theta_k^{(i)} = \frac{2\pi k}{K_i} \in [0, 2\pi ) , \quad k = 0,1,\cdots,K_i-1.\\
\end{aligned}
\end{equation}
The radius $R_i$ is selected according to the robust design criterion in \cite{walk2019mocz}, which maximizes the minimum separation between neighboring zero locations and thereby improves noise robustness. For each submessage, the parameter $K_i$ determines the angular resolution of the corresponding zero constellation, while the bits of $\mathbf m^{(i)}$ select the inner or outer radius at each angular position. Unlike conventional BMOCZ, which employs a single zero constellation of size $K$, the proposed MR-BMOCZ constructs the transmitted waveform by jointly combining $I$ independent zero constellations of sizes $\{{K_i}\}_{i=1}^{I}$. This additional degree of freedom enables the transmitted information to be distributed across multiple zero constellations with different angular resolutions. Hence, the overall zero set can be shaped more flexibly than in conventional BMOCZ while preserving the fundamental BMOCZ encoding principle.

As illustrated in Fig.~\ref{blockdiagram}, each submessage is independently mapped to a polynomial through the corresponding zero constellation. Specifically, the polynomial associated with the $i$-th submessage is given by
\begin{equation}
    X_i(z) = x_{K_i}^{(i)}\prod_{k=0}^{K_i-1}\bigl(z-\alpha_{k}^{(i)}\bigr).
\end{equation}
The transmitted polynomial can be defined as the product of multiple \ac{BMOCZ} polynomials, represented as
\begin{equation} \label{multiplacationX}
    X(z) =x_{K} \prod_{i=1}^{I} \prod_{k=0}^{K_i-1}\bigl(z-\alpha_{k}^{(i)}\bigr),
\end{equation}
with $x_{K} = \prod_{i=1}^{I}x_{K_i}^{(i)}$. It follows that the transmitted sequence $\mathbf{x}=\{x[n]\}_{n=0}^{K}$ can be regarded as the linear convolution of multiple Huffman sequences, expressed as
\begin{equation} \label{eqautocorr}
    \mathbf{x} = \mathbf{x}^{(1)} \ast \mathbf{x}^{(2)} \ast \cdots \ast \mathbf{x}^{(i)}\ast \cdots \ast\mathbf{x}^{(I)}.
\end{equation}
This convolutional representation highlights that the overall transmitted sequence is generated by combining multiple Huffman sequences associated with different submessages. In the polynomial domain, this corresponds to the union of the zero sets of the individual polynomials, resulting in a structured multi-radius zero constellation in the complex plane.

\subsection{Sub-polynomial angle design}
Although the multi-radius construction in \eqref{multiplacationX} guarantees uniform spacing within each individual zero set, it does not necessarily maximize the separation between zeros originating from different submessages. As a result, zeros belonging to distinct radii may become closely spaced, which reduces the minimum inter-set distance. This effect can introduce inter-set interference, increase noise sensitivity, and ultimately degrade detection performance\cite{walk2020mocz}. To mitigate this issue, we introduce a controlled phase offset applied to each zero set. Specifically, the zeros associated with submessage $\mathbf{m}^{(i)}$ are rotated by a fixed phase $\Phi_i^{(0)}$, resulting in a global angular shift of the corresponding constellation. The phase offsets are selected to maximize the minimum angular separation between zeros originating from different constellations. We define the relative phase offset between two grids as $\Delta\Phi_{i,j} \triangleq \Phi_j^{(0)}-\Phi_i^{(0)} \pmod{2\pi}$. The conditions under which inter-set interference occurs between different zero sets, as well as the optimal choice of $\Phi_i^{(0)}$ that maximizes their separation, are characterized by the following lemma.

\textbf{\textit{Lemma 1 (Angular Spacing and Optimal Phase Offset).}}
Considering the case of two radius groups ($I=2$) containing $K_i$ and $K_j$ uniformly spaced angular positions. The minimum angular separation between zeros belonging to different groups is maximized when the relative phase offset is chosen as
\begin{equation} 
\Delta\Phi_{i,j}^{\star}
=
\frac{\pi}{G_{i,j}},
\qquad
G_{i,j}=\operatorname{lcm}(K_i,K_j),
\label{eq:optimalphase}
\end{equation}
where $\mathrm{lcm}(\cdot,\cdot)$ denotes the least common multiple. This choice shifts the $j$-th angular grid relative to the $i$-th angular grid, improving the  zeros' geometric separation in the complex plane. The angular separation for $I>2$ follows recursively by applying the same procedure to each additional radius pair

\textbf{Proof:} Please refer to Appendix A. \hfill$\blacksquare$

\begin{figure}[t!]
\centering
\resizebox{0.9\columnwidth}{!}{
\includegraphics{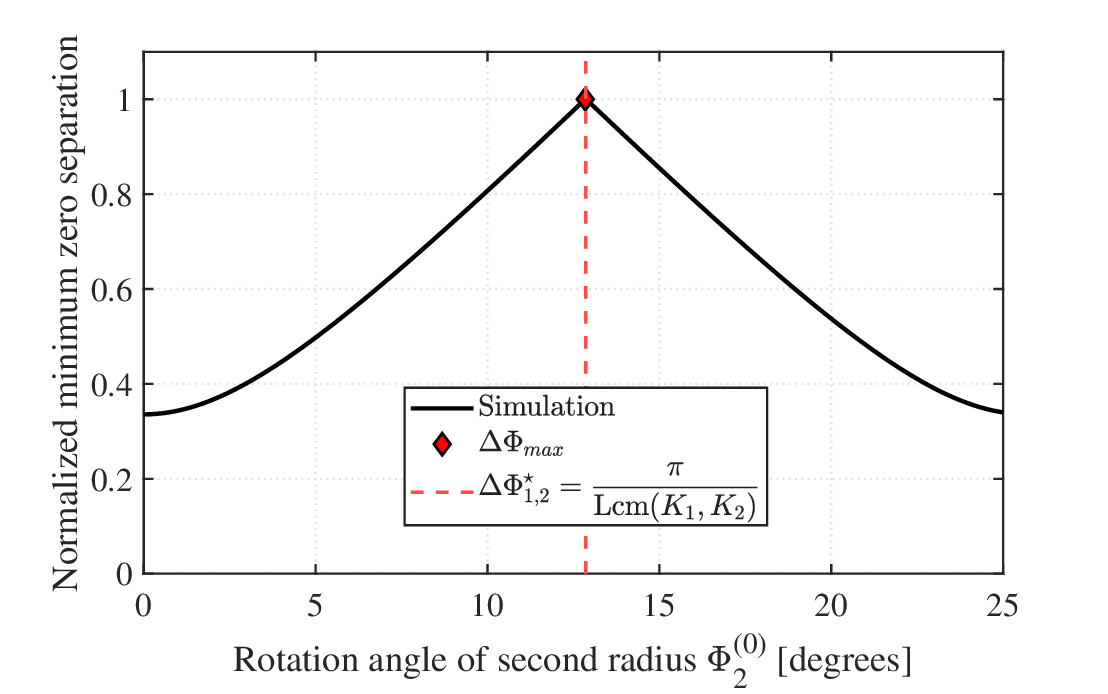}}
\caption{Normalized minimum zero separation versus rotation angle of the second-radius zero set. The optimal rotation $\Delta\Phi_{1,2}^{\star}$ maximizes the minimum distance between zeros.}
\label{maxPhase}
\end{figure}

\begin{figure}[t!]
\centering
\begin{subfigure}[b]{0.9\columnwidth}
    \centering
    \includegraphics[
   width=1\linewidth,
        trim=2.2cm 0.1cm 2.2cm 0.8cm,
        clip
    ]{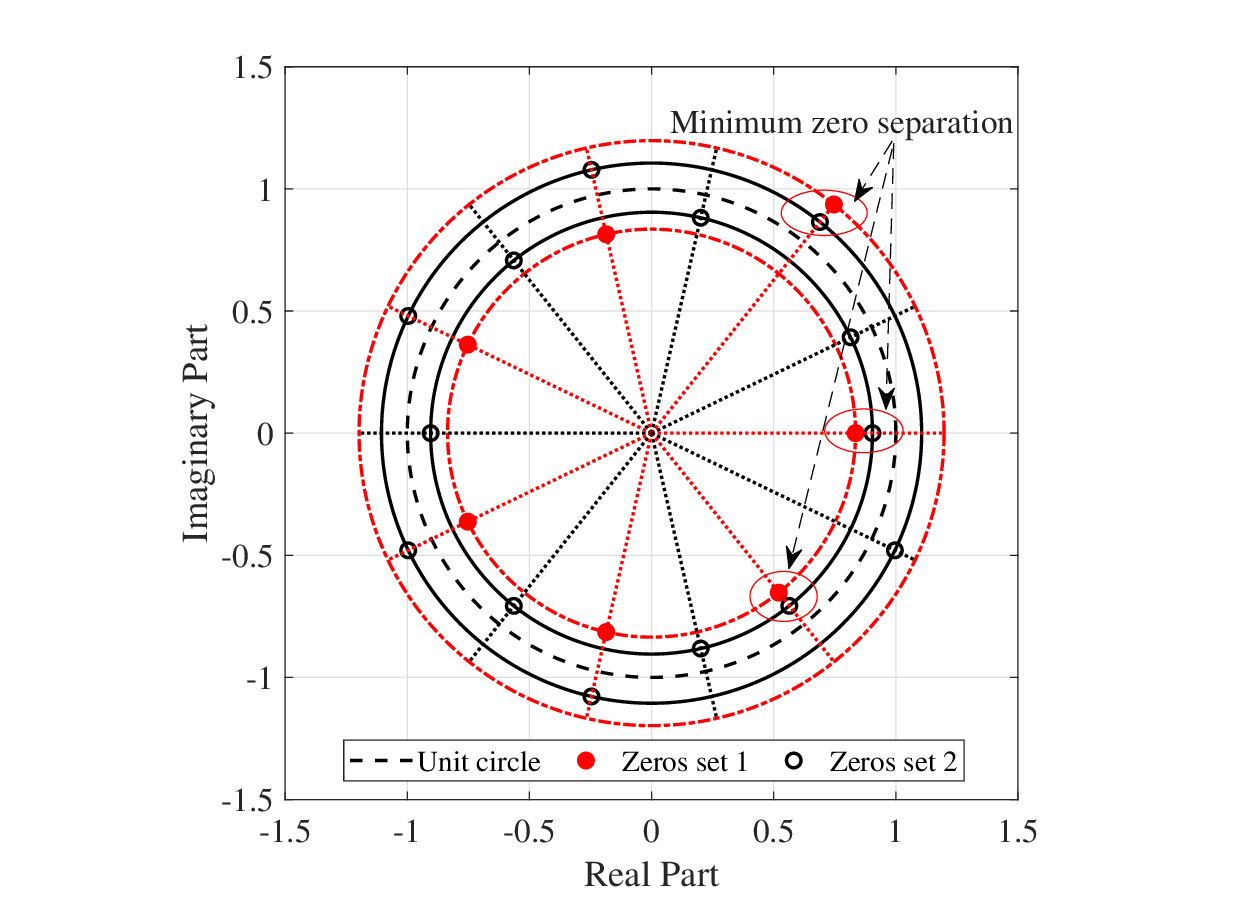}
    \caption{}
\end{subfigure}
\begin{subfigure}[b]{0.9\columnwidth}
    \centering
    \includegraphics[        width=1\linewidth,
        trim=2.2cm 0.1cm 2.2cm 0.8cm,
        clip
    ]{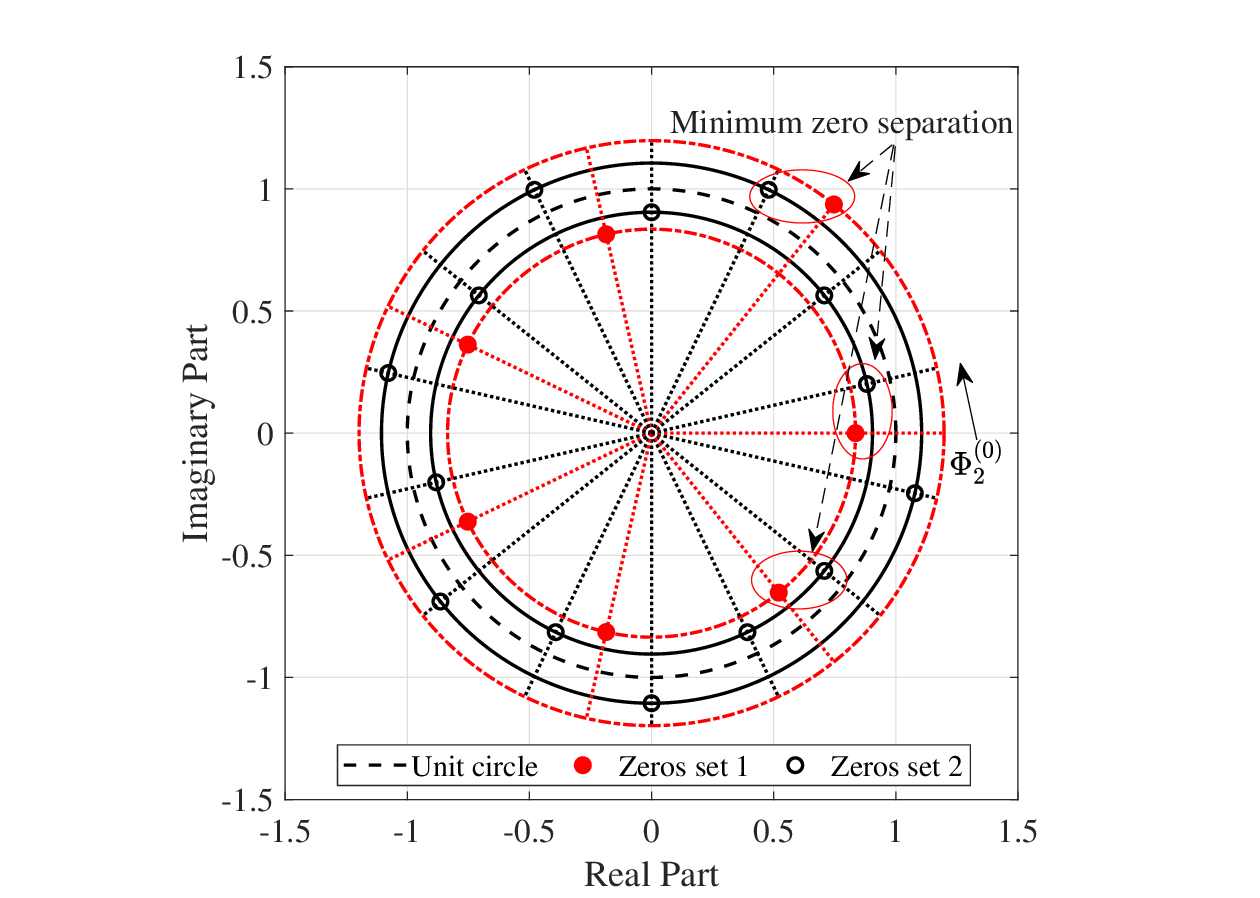}
    \caption{}
\end{subfigure}
\caption{Basic implementation of two radii MR-BMOCZ with $K_1=7$, $K_2=14$ (a) before angle optimization. (b) with a rotation angle of $\Delta\Phi_{1,2}^{\star} =
\frac{\pi}{14}$.}
\label{fig:two_radius_mrbmocz}
\end{figure}

Lemma~1 establishes that the zeros associated with different radius groups become maximally separated in the angular domain by selecting the relative phase offsets according to \eqref{eq:optimalphase}. This ensures that the minimum angular separation between zeros belonging to different radius groups is maximized. This is illustrated in Fig.~\ref{maxPhase}, where the normalized minimum zero separation attains its maximum value at the optimal phase rotation $\Delta\Phi_{1,2}^{\star} = \Phi_2^{(0)}-\Phi_1^{(0)}$, where for $I=2$ we have $\Phi_1^{(0)}=0$, thereby validating the theoretical result.
Importantly, the proposed angular optimization does not affect the radial placement of the zeros, which remains identical to that of the single-radius \ac{BMOCZ} framework. Consequently, the optimal radial separation of each zero set is preserved \cite{walk2019mocz}, while the additional angular phase shifts provide separation between the submessages. Fig.~\ref{fig:two_radius_mrbmocz} illustrates a two-radius realization of the proposed \ac{MR-BMOCZ} framework.

\subsection{Receiver-Side Processing}
At the receiver side, decoding is performed using a generalized \ac{DiZeT} structure. It is assumed that both the transmitter and the receiver have prior knowledge of the number of radius groups $I$ as well as $\{K_i\}_{i=1}^{I}$, which uniquely determine the angular shifts associated with the submessages $\mathbf{m}^{(i)}$ according to Lemma~1. Under this assumption, symbol detection can be carried out independently for each radius group. The estimated symbol corresponding to the $k$-th angular position of the $i$-th submessage is given by
\begin{equation}
\begin{aligned}
\hat{m}^{(i)}_{k}
&=
\frac{1}{2}\left\{
1+\mathrm{sign}\left(
R_i^{-\frac{K_i}{2}}
\left|Y\left(R_i e^{j\theta_k^{(i)}}\right)\right|
\right.\right. \\
&\left.\left.
-
R_i^{\frac{K_i}{2}}
\left|Y\left(R_i^{-1} e^{j\theta_k^{(i)}}\right)\right|
\right)
\right\}.
\end{aligned}
\end{equation}
where $\mathrm{sign}(\cdot)$ denotes the sign function. This decision rule compares the magnitudes of the received $z$-transform evaluated at reciprocal points on the complex plane, scaled to compensate for the polynomial degree. The resulting sign therefore performs a binary radial test that maps the outcome directly to the transmitted symbol $\hat{m}^{(i)}_{k}\in\{0,1\}$. By applying this test independently for each angular index and radius group, the decoder recovers all submessages $\mathbf{\hat{m}}^{(i)}$ without requiring explicit channel estimation. Finally, the original message is reconstructed by concatenating the estimated symbols across all radius groups as
\begin{equation}
    \hat{\mathbf{m}} = [\mathbf{\hat{m}}^{(1)}, \cdots, \mathbf{\hat{m}}^{(I)}].
\end{equation}
The corresponding receiver structure is illustrated in Fig.~\ref{blockdiagram}.


\section{AUTOCORRELATION DIVERSITY AND SYNCHRONIZATION}
\label{sec:sync}
The proposed \ac{MR-BMOCZ} waveform exhibits an autocorrelation structure that differs fundamentally from that of conventional Huffman-\ac{BMOCZ}. In the following, we analyze the resulting autocorrelation structure and demonstrate how the generated peaks can be exploited for both \ac{TO} and \ac{CFO} estimation.

\subsection{Autocorrelation Structure}
In conventional Huffman-\ac{BMOCZ}, the constellation set defined in \eqref{map} provides a near-impulsive \ac{AACF} structure \eqref{impulsive}. This plays a fundamental role in blind synchronization, as the dominant autocorrelation peaks can be exploited for timing acquisition and synchronization-related signal processing \cite{walk2020mocz}. Since the proposed \ac{MR-BMOCZ} waveform is constructed as the convolution of multiple \ac{BMOCZ} sub-sequences, its autocorrelation no longer consists of only the three deterministic Huffman peaks. Instead, the convolution of the individual Huffman autocorrelations generates additional deterministic peaks whose indices are determined by the selected values of ${K_i}$. These peaks provide multiple synchronization references in the autocorrelation domain that are exploited to improve synchronization performance. 

\begin{figure}[t!]
\centering
\resizebox{0.9\columnwidth}{!}{
\includegraphics{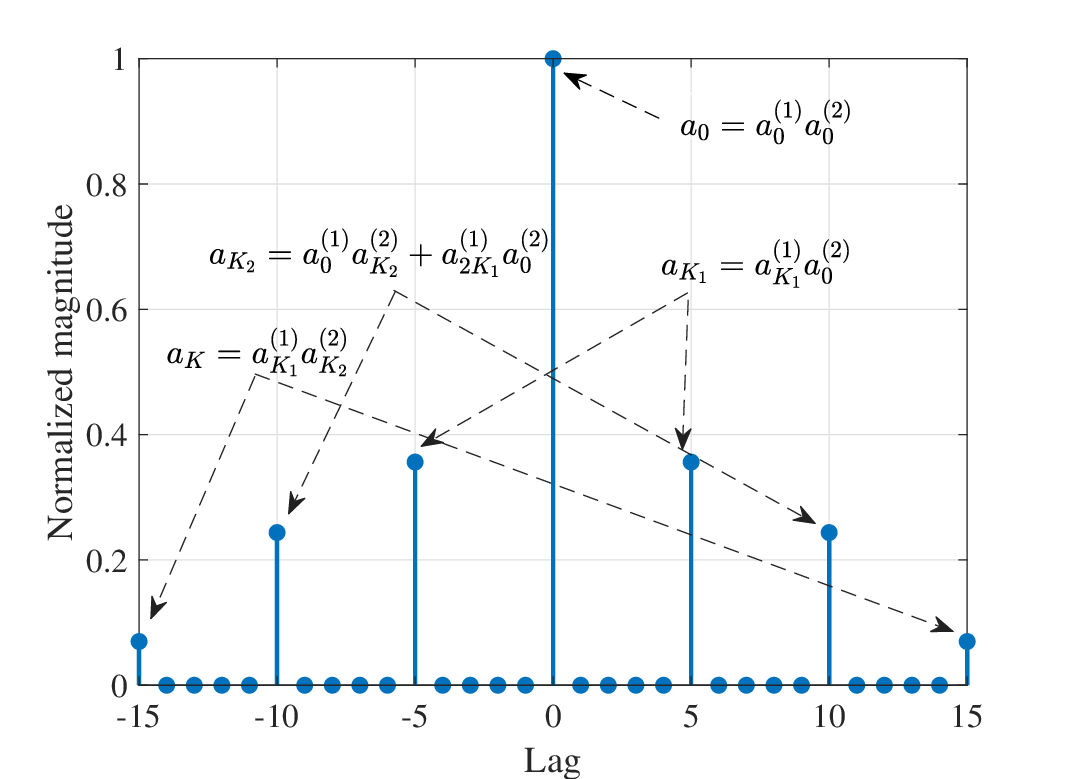}}
\caption{Autocorrelation of the \ac{MR-BMOCZ} sequence for $K_2 = 2K_1 (K_1 = 7 \;,\; K_2 =14)$,where the peak at $n=0$ along with six side peaks are all used for better \ac{TO} estimation.}
\label{bracketcorr7}
\end{figure}

For the $i$-th sub-polynomial, the \ac{AACF} is defined as
\begin{equation}
a_{\mathrm{MR}}^{(i)} = \mathbf{x}^{(i)} \ast \overline{(\mathbf{x}^{(i)})^{-}}
= (-\eta^{(i)},0_{K_i},1,0_{K_i},-\eta^{(i)}),
\label{corrpolyi}
\end{equation}
where $\eta^{(i)} = \bigl(R_i^{K_i}+R_i^{-K_i}\bigr)^{-1}$. Therefore, each sub-sequence contains a dominant center peak of unit amplitude, associated with two symmetric side peaks of magnitude $\eta^{(i)}$. The autocorrelation of the overall MR-BMOCZ waveform is then obtained by combining the autocorrelations of all constituent sub-sequences. Since the transmitted sequence is generated according to \eqref{multiplacationX}, the corresponding \ac{AACF} is expressed as
\begin{equation}
\begin{aligned}
a_{\mathrm{MR}} &= \left(x^{(1)} * \cdots * x^{(I)}\right)
   * \overline{\left(x^{(1)} * \cdots * x^{(I)}\right)^{-}}  \\
&= \left(x^{(1)} * \overline{(x^{(1)})^{-}}\right)
 * \cdots *
   \left(x^{(I)} * \overline{(x^{(I)})^{-}}\right) \\
&= a^{(1)} * a^{(2)} * \cdots * a^{(I)} .
\end{aligned}
\label{eq:corr22}
\end{equation}
The result in \eqref{eq:corr22} establishes that the \ac{AACF} of the proposed \ac{MR-BMOCZ} waveform is obtained by convolving multiple Huffman sequences. Thus, the resulting \ac{AACF} contains more deterministic peaks than conventional \ac{BMOCZ}, thereby introducing additional autocorrelation diversity. Note that the \ac{AACF} can also be represented in the z-domain as
\begin{equation}
    A(z) = z^{K}X(z)\overline{X\left(1/\bar z\right)},
\end{equation}
Using the factorized \ac{MR-BMOCZ} structure in \eqref{multiplacationX}, we obtain
\begin{equation}
\begin{aligned}
A(z) &= x_K \bar{x}_0\prod_{i=1}^{I}\prod_{k=0}^{K_i-1}(z-\alpha_k^{(i)}) \prod_{i=1}^{I}\prod_{k=0}^{K_i-1}(z-\frac{1}{\bar{\alpha}_k^{(i)}}),\\ &=x_K\bar{x}_0 \prod_{i=1}^{I}X_i(z)\overline{X_i\left(\frac{1}{\bar z}\right)}=\prod_{i=1}^{I}A^{(i)}(z),
\end{aligned}
\label{eq:AACFzMR}
\end{equation}
where $A^{(i)}(z) = z^{K_i}X_i(z)\overline{X_i\left(1/\bar z\right)}$
denotes the z-domain \ac{AACF} of the $i$-th Huffman-\ac{BMOCZ} sub-polynomial.
\begin{table}[t!]
\begin{center}
\caption{Nonzero autocorrelation components of the two-radius MR-BMOCZ structure.}
\label{tab:autocorr}
\begin{tabular}{|l|c|c|}
\hline
Index & $K_1<K_2<2K_1$ & $K_2=2K_1$ \\
\hline\hline
$0$ &
$a^{(1)}_0 a^{(2)}_0$ &
$a^{(1)}_0 a^{(2)}_0$ \\

$K_1$ &
$a^{(1)}_{K_1} a^{(2)}_0$ &
$a^{(1)}_{K_1} a^{(2)}_0$ \\

$K_2$ &
$a^{(1)}_0 a^{(2)}_{K_2}$ &
$a^{(1)}_0 a^{(2)}_{K_2}+a^{(1)}_{2K_1} a^{(2)}_0$ \\

$2K_1$ &
$a^{(1)}_{2K_1} a^{(2)}_0$ &
\textit{merged with } $a_{K_2}$ \\

$K_1+K_2$ &
$a^{(1)}_{K_1} a^{(2)}_{K_2}$ &
$a^{(1)}_{K_1} a^{(2)}_{K_2}$ \\

$2K_1+K_2$ &
$a^{(1)}_0 a^{(2)}_{K_2}$ &
$a^{(1)}_0 a^{(2)}_{K_2}+a^{(1)}_{2K_1} a^{(2)}_0$ \\

$K_1+2K_2$ &
$a^{(1)}_{K_1} a^{(2)}_0$ &
$a^{(1)}_{K_1} a^{(2)}_0$ \\

$2K$ &
$a^{(1)}_0 a^{(2)}_0$ &
$a^{(1)}_0 a^{(2)}_0$ \\
\hline
\end{tabular}
\end{center}
\end{table}
To illustrate this behavior, consider the two-radius case $I=2$. Depending on the relationship between $K_1$ and $K_2$, two different peak structures can arise. Table~\ref{tab:autocorr} summarizes the \ac{AACF} peak locations, where $a^{(i)}_k$ denotes the $k$-th element of the \ac{AACF} for the $i$-th sub-polynomial as given in \eqref{corrpolyi}. When $K_1<K_2<2K_1$, the generated peaks are generally non-uniformly spaced. In contrast, when $K_2=2K_1$, several peaks overlap, resulting in a more regular and symmetric autocorrelation structure, as illustrated in Fig.~\ref{bracketcorr7}. Accordingly, the remainder of this work considers the two-radius case $I=2$ with $K_2=2K_1$, which produces a symmetric and regularly spaced autocorrelation signature. The resulting predictable peak structure directly supports the proposed timing template and synchronization analysis.

An additional advantage of the proposed multi-radius construction is the increase in deterministic autocorrelation energy. For conventional BMOCZ, the normalized \ac{AACF} in \eqref{impulsive} yields a total autocorrelation energy as
\begin{equation}
E_{\mathrm{B}} = \sum_n |a_{\mathrm{B}}[n]|^2 = 1+2\eta^2.
\end{equation}
For the proposed MR-BMOCZ waveform, the AACF is obtained by convolving the individual Huffman AACFs according to \eqref{eq:corr22}. For the two-radius case ($I=2$) we have $a_{\mathrm{MR}} = a^{(1)} \ast a^{(2)}$.
Since each autocorrelation contains three nonzero peaks, their convolution generates seven deterministic peak combinations. Accordingly, the total deterministic autocorrelation energy becomes
\begin{equation} \label{Energy1}
E_{\mathrm{MR}} = 1+2\eta_1^2+2\eta_2^2+4\eta_1^2\eta_2^2.
\end{equation}
More generally, for $I$ radius groups we have
\begin{equation}
\label{Energy2}
E_{\mathrm{MR}} = \prod_{i=1}^{I}\left(1+2\eta_i^2\right),
\end{equation}
From \eqref{Energy1} and \eqref{Energy2}, it can be observed that $E_{\mathrm{MR}} \ge E_B$. Since synchronization metrics are directly related to the available \ac{AACF} energy, the additional side peaks provide more timing observations and improve the reliability of synchronization. This property forms the foundation of the \ac{TO} and \ac{CFO} methods developed in the following subsections.
\subsection{Proposed Timing Offset Estimation}
The synchronization advantage of the proposed \ac{MR-BMOCZ} waveform does not only originate from a larger autocorrelation energy, but from the fact that this energy is distributed over a deterministic radius-induced peak lattice. For the $i$-th radius group, the AACF is given by
\begin{equation}
a^{(i)}
=
\left(
-\eta^{(i)},
0_{K_i},
1,
0_{K_i},
-\eta^{(i)}
\right),
\end{equation}
whose index is $\mathcal{S}_i=\{-K_i,0,K_i\}$, and therefore the \ac{MR-BMOCZ} autocorrelation index becomes the Minkowski sum\cite{weibel2007minkowski}
\begin{equation}
    \mathcal{S}_{\mathrm{MR}}
    =
    \mathcal{S}_1 \oplus \mathcal{S}_2 \oplus \cdots \oplus \mathcal{S}_I .
\end{equation}
Hence, each deterministic timing peak is located at
\begin{equation} \kappa=\sum_{i=1}^{I}\nu_iK_i, \quad\nu_i\in\{-1,0,1\}.
\end{equation}
The timing signature is no longer described by only the two boundary side peaks of conventional Huffman-\ac{BMOCZ}, but by a set of predictable peaks whose locations are controlled by the selected radius partition $\{K_i\}_{i=1}^{I}$. The received signal can be expressed as
\begin{equation}
\tilde{y}[n] = e^{jn\varepsilon} (x*h)[n-\tau] + \tilde{w}[n],
\end{equation}
where $\tau$ denotes the unknown integer-valued symbol arrival time and $\varepsilon$ represents the \ac{CFO}. For clarity, the \ac{TO} estimation problem is first considered in the absence of \ac{CFO}, i.e., $\varepsilon=0$. Subsequently, the more general scenario involving simultaneous timing and frequency offsets is investigated for arbitrary values of $\varepsilon$. Since the dominant autocorrelation peaks of the transmitted waveform occur at the known positions in $\mathcal{S}_{\mathrm{MR}}$, timing acquisition can be performed by correlating the received waveform with the complete MR peak template rather than with a single peak. Accordingly, we define the MR timing template as
\begin{equation}
    g_{\mathrm{MR}}[n]
    =
    \sum_{\kappa\in\mathcal{S}_{\mathrm{MR}}}
    \gamma_{\kappa} a_{\mathrm{MR}}^{*}[\kappa]\delta[n-\kappa],
\end{equation}
where $a_{\mathrm{MR}}[\kappa]$ denotes the deterministic \ac{MR-BMOCZ} autocorrelation coefficient and $\gamma_{\kappa}\geq 0$ is a weighting coefficient introduced to suppress the influence of weak autocorrelation peaks, defined as
\begin{equation}
\gamma_{\kappa}
=\frac{|a_{\mathrm{MR}}[\kappa]|^2} {\sum_{q\in\mathcal{S}_{\mathrm{MR}}}|a_{\mathrm{MR}}[q]|^2}.
\end{equation}
The proposed timing metric is then given by
\begin{equation}
    \Lambda_{\mathrm{MR}}[n] =
    |(g_{MR} \ast r_{\tilde{y}})[n]|^2    =\left|\sum_{\kappa\in\mathcal{S}_{\mathrm{MR}}}\gamma_{\kappa}a_{\mathrm{MR}}^{*}[\kappa] r_{\tilde{y}}[n-\kappa]\right|^2,
\end{equation}
where $r_{\tilde{y}}[\cdot]$ denotes the local correlation output obtained from the received samples. A low-complexity implementation can be obtained by using the four-tap MR Huffman bracket
\begin{equation}
    \breve{\varkappa}_{\psi}
    =
    e^{-j\psi}\delta[n]
    +
    \delta[n-K_1]
    +
    e^{-j\psi}\delta[n-K_2]
    +
    \delta[n-K].
\end{equation}
The corresponding output is
\begin{equation}
\begin{aligned}
(\breve{\varkappa}_{\psi}*\tilde{y})[n]
=&
e^{-j\psi}\tilde{y}[n]
+\tilde{y}[n-K_1]  \\
&+
e^{-j\psi}\tilde{y}[n-K_2]
+\tilde{y}[n-K].
\end{aligned}
\end{equation}
Unlike the conventional two-tap Huffman bracket, this bracket is matched to the \ac{AACF} structure of \ac{MR-BMOCZ}. The corresponding timing metric is
\begin{equation}
    \Lambda_{\mathrm{MR}}[n]
    =
    \left|
    (\breve{\varkappa}_{\psi}*\tilde{y})[n]
    \right|^2 .
\end{equation}

At the correct timing hypothesis, the radius-induced peaks align simultaneously with the template positions, and their contributions add coherently. Therefore, the \ac{TO} estimate is obtained as
\begin{equation}
    \hat{\tau}
    =
    \arg\max_{n}\Lambda_{\mathrm{MR}}[n].
\end{equation}

In the presence of an arbitrary \ac{CFO}, several shifted brackets are averaged to reduce sensitivity to the unknown phase relation between the bracket components\cite{walk2020mocz}. Let $\psi_p=2\pi p/P$, for  $p=0,\ldots,P-1$ denote the set of $P$ uniformly spaced phase offsets associated with the different brackets. The final MR timing metric is
\begin{equation}
    \Gamma_{\mathrm{MR}}[n]
    =
    \frac{1}{P}\sum_{p=0}^{P-1}
    \left|
    (\breve{\varkappa}_{\psi_p}*\tilde{y})[n]
    \right|^2 ,
\end{equation}
and the timing estimate will be given as
\begin{equation}
    \hat{\tau}
    =
    \arg\max_n \Gamma_{\mathrm{MR}}[n].
\end{equation}

\subsection{Proposed Carrier Frequency Offset Estimation}

After timing correction, the received signal in the presence of \ac{CFO} can be written as
\begin{equation}
    \tilde{y}[n]
    =
    e^{j\varepsilon n}(x*h)[n]
    +
    \tilde{w}[n],
\end{equation}
where $\varepsilon=2\pi\Delta fT_s$ is the normalized \ac{CFO}. In the polynomial domain, this impairment has a particularly useful interpretation. The received signal can be written as
\begin{equation}
    \tilde{Y}(z)=\sum_n \tilde{y}[n]z^n,
\end{equation}
Therefore, all signal and channel zeros are rotated by the same angle $\alpha_{k}^{(i)} \longrightarrow \alpha_{k}^{(i)}e^{-j\varepsilon}.$ This common angular drift is harmful for \ac{DiZeT} detection because the receiver evaluates the received polynomial at the nominal MR constellation angles. However, it also provides an estimation opportunity. For \ac{MR-BMOCZ}, the \ac{CFO} can jointly align all radius groups with their corresponding angular grids. Such rotations produce indistinguishable constellations and therefore represent intrinsic \ac{CFO} ambiguities. Therefore, before developing the estimator, we first identify the set of integer-\ac{CFO} phase rotations that leave the multi-radius zero constellation invariant.

\begin{figure}[t!]
\centering
\resizebox{0.9\columnwidth}{!}{
\includegraphics{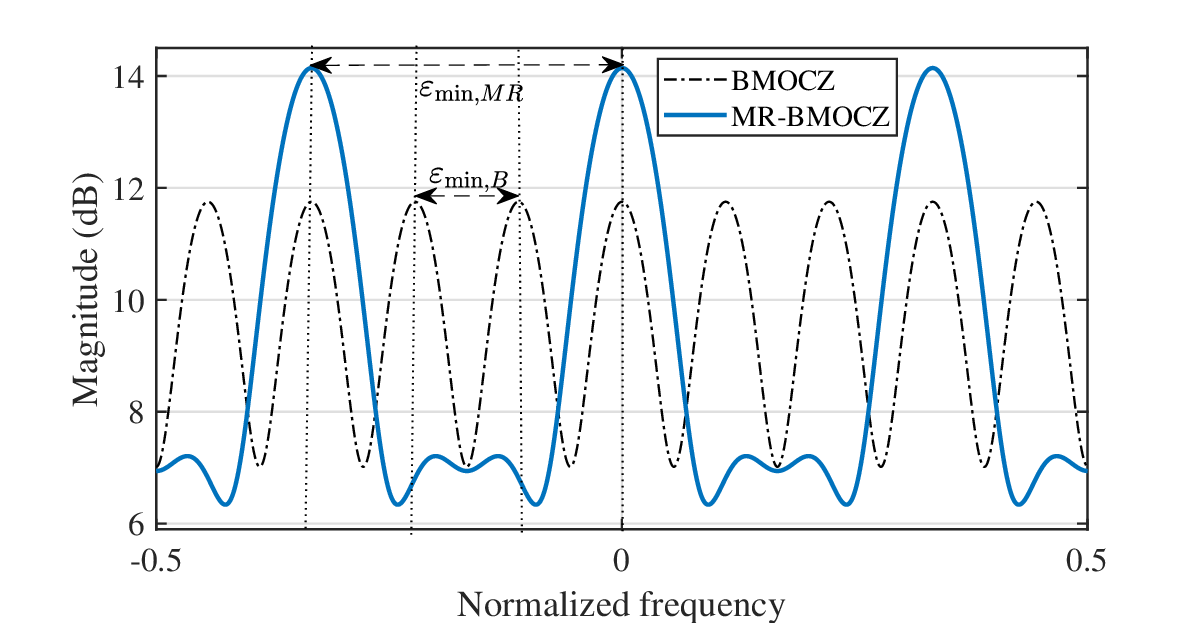}}
\caption{Power spectral density of the proposed \ac{MR-BMOCZ} compared to conventional \ac{BMOCZ}}
\label{Spectrumall}
\end{figure}

\textbf{\textit{Lemma 2 (CFO Unambiguous Condition).}}
For a multi-radius constellation composed of $I$ angular grids with
cardinalities $K_1,\ldots,K_I$, the constellation remains invariant
under a \ac{CFO}-induced phase rotation if and only if
\begin{equation}
    \varepsilon
    =
    \frac{2\pi q}
    {\gcd(K_1,\ldots,K_I)},
    \qquad
    q \in \mathbb{Z}.
\end{equation}
In particular, the minimum nonzero \ac{CFO} phase rotation $\varepsilon_{min}$ that leaves the constellation invariant corresponds to $q=1$.
\textbf{Proof:} Please refer to the proof in Appendix~B.\hfill$\blacksquare$

Lemma~2 characterizes the rotational symmetry of the proposed multi-radius zero constellation. This periodicity is determined by the smallest nonzero \ac{CFO}-induced phase rotation for which the constellation maps onto itself. As a result, the parameter $\varepsilon_{\min}$ defines the fundamental rotational period of the constellation and directly specifies the range over which \ac{CFO} can be uniquely identified from the received zero pattern. Under the condition $\varepsilon < \varepsilon_{\min}$\footnote{This assumption is consistent with \cite{huggins2025fourier}, where the maximum CFO is reported as $\pm120$~kHz at a carrier frequency of $2.4$~GHz, corresponding to a normalized range of approximately $[-0.006,0.006]$. Therefore, it is reasonable to assume $|\varepsilon| < \frac{1}{2}\varepsilon_{\min}$.}, the dominant spectral peak present in Fig.~\ref{Spectrumall} generated by the \ac{MR-BMOCZ} structure can be exploited for \ac{CFO} estimation. Note that the proposed \ac{MR-BMOCZ} scheme provides a wider unambiguous \ac{CFO} estimation range $\varepsilon_{min,MR}$ than that of conventional \ac{BMOCZ}, denoted by $\varepsilon_{min,B}$, allowing larger frequency offsets to be uniquely identified from the received zero constellation.

For a candidate CFO value $\varepsilon_c$, the compensated polynomial is $\tilde{Y}_{\varepsilon_c}(z) = \tilde{Y}(e^{-j\varepsilon_c}z)$. The CFO can then be estimated by searching for the candidate that maximizes the multi-radius radial contrast used by the generalized \ac{DiZeT} detector. Specifically, we define
\begin{equation}
\begin{aligned}
\mathcal{J}_{\mathrm{MR}}(\varepsilon_{c})
=
\sum_{k=0}^{K_{min}-1}
\Bigg|
&
R_{min}^{-\frac{K_{min}}{2}}
\left|
\tilde{Y}_{\varepsilon_c}
\left(
R_{min} e^{j\theta_{k}}
\right)
\right|   \\
&-
R_{min}^{\frac{K_{min}}{2}}
\left|
\tilde{Y}_{\varepsilon_c}
\left(
R_{min}^{-1} e^{j\theta_{k}}
\right)
\right|
\Bigg|,
\end{aligned}
\end{equation}
where $K_{min} \triangleq  \min_{i\in\{1,\ldots,I\}} K_i$, $R_{min}$ and $\theta_{k}$ denote the radius and the angular grid associated with the zero group $K_{min}$, respectively. The proposed \ac{CFO} estimate is then obtained as
\begin{equation}
    \hat{\varepsilon}
    =
    \arg\max_{\varepsilon_c\in\Omega}
    \mathcal{J}_{\mathrm{MR}}(\varepsilon_c),
\end{equation}
where $\Omega$ denotes the considered CFO search grid. The objective in the above equation is constrained by several angular grids. An incorrect \ac{CFO} candidate that aligns the zeros associated with a particular radius group fails to align the zeros of the remaining groups. Hence, the multi-radius construction sharpens the CFO metric and reduces ambiguous frequency hypotheses.
\section{PERFORMANCE ANALYSIS} \label{Performance}
In this section, we analyze the \ac{PAPR} behavior, noise robustness and computational complexity of the proposed \ac{MR-BMOCZ} construction compared with conventional \ac{BMOCZ}.
\subsection{\ac{PAPR} Analysis}
The \ac{PAPR} of a given transmitted vector $\mathbf{x} =[x_0,x_1,\ldots,x_K]^T$ is defined as
\begin{equation}
\mathrm{PAPR}=\frac{(K+1)\max_n |x_n|^2}{\|\mathbf{x}\|_2^2},
\label{eq:PAPR}
\end{equation} From \eqref{multiplacationX} and \eqref{eqautocorr}, the proposed \ac{MR-BMOCZ} polynomial is constructed as the product of $I$ \ac{BMOCZ} sub-polynomials $X(z)=\prod_{i=1}^{I}X_i(z)$. This distributed construction modifies the coefficient distribution and directly affects the energy concentration of the resulting waveform. Using Parseval's theorem, the signal energy is given by
\begin{equation} \label{sigEnergy}
E= \frac{1}{2\pi}
\int_{0}^{2\pi}
\prod_{i=1}^{I}
\left|X_i\left(e^{j\omega}\right)\right|^2 d\omega .
\end{equation}
For each sub-polynomial, the edge coefficient is given as
\begin{equation}
|x_0^{(i)}|^2
=
|x_{K_i}^{(i)}\prod_{k=0}^{K_i-1}\alpha_k^{(i)}|^2
=
|x_{K_i}^{(i)}|^2
R_i^{2(n_{1,i}-n_{0,i})},
\end{equation}
where $n_{1,i}$ and $n_{0,i}$ denote the number of bits mapped to the outer and inner radii, respectively. Since the overall \ac{MR-BMOCZ} sequence is obtained by convolution, the extreme coefficients satisfy $x_0=\prod_{i=1}^{I}x_0^{(i)}$ and $x_K=\prod_{i=1}^{I}x_{K_i}^{(i)}$. Therefore, the magnitude of the edge coefficients is strongly influenced by the product of the radii associated with each sub-polynomial. In the worst-case configuration, where all bits are mapped onto the outer radii, the edge coefficient scales as
\begin{equation}
|x_0|_{\max}^2
=
\prod_{i=1}^{I}
|x_{K_i}^{(i)}|^2 R_i^{2K_i}.
\label{PAPRfinal}
\end{equation}

Equation \eqref{PAPRfinal} characterizes the coefficient growth of the proposed \ac{MR-BMOCZ} waveform. Unlike conventional \ac{BMOCZ}, where the polynomial structure is concentrated in a single high-order polynomial $R^{2K}$, the proposed approach distributes the signal generation across multiple lower-order sub-polynomials $\prod_{i=1}^{I}R_i^{2K_i}$. As a result, the energy is spread across the convolutional coefficient structure, reducing the concentration of power in a small number of samples. This behavior results in a lower tendency toward large-amplitude peaks and consequently yields a reduced \ac{PAPR} compared with conventional \ac{BMOCZ}.
\subsection{Sensitivity Analysis}
The detection performance of \ac{BMOCZ}-based signaling is fundamentally governed by the sensitivity of the information-bearing zeros to additive noise. Since the DiZeT detector operates by evaluating the received polynomial at the expected zero locations, any perturbation of the transmitted zeros directly affects the reliability of the binary decisions. For analytical tractability, we consider the \ac{AWGN} case with variance $N_0$ and neglect synchronization errors. Let $\alpha_k$ denote a transmitted zero of $X(z)$. In the presence of noise, the zero is displaced to $\hat{\alpha}_k=\alpha_k+\Delta\alpha_k$, where $\Delta\alpha_k$ is modeled as a zero-mean Gaussian random variable. The mean-squared root perturbation can be approximated as
\begin{equation}
\mathbb{E}\{|\Delta\alpha_k|^2\}
\approx
\frac{
N_0
\sum_{n=0}^{K}
|\alpha_k|^{2n}
}
{|X'(\alpha_k)|^2},
\label{eq:rootPert}
\end{equation}
Equation \eqref{eq:rootPert} shows that the robustness of a zero against noise is directly determined by the derivative magnitude of the transmitted polynomial at that location. Larger zero separations increase $|X'(\alpha_k)|$, reduce the expected root displacement, and improve detection reliability. For the proposed \ac{MR-BMOCZ} scheme, the situation is fundamentally different. From \eqref{multiplacationX}, we have $X(z)=X_1(z)X_2(z)$, and let $\alpha_k^{(1)}$ denote a zero belonging to the first radius group. Since
\begin{equation}
X_1(\alpha_k^{(1)})=0,
\qquad
X_2(\alpha_k^{(1)})\neq 0,
\end{equation}
the derivative of the overall polynomial becomes
\begin{equation}
X'(\alpha_k^{(1)})
=
X_1'(\alpha_k^{(1)})
X_2(\alpha_k^{(1)}).
\end{equation}
Defining $\Psi_k^{(1)} \triangleq X_2(\alpha_k^{(1)})$, the derivative magnitude can be expressed as
\begin{equation}
|X'(\alpha_k^{(1)})|
=
|X_1'(\alpha_k^{(1)})|
\Psi_k^{(1)}|.
\label{eq:MRderivative}
\end{equation}
Using the factorized form of $X_2(z)$,
\begin{equation}
\Psi_k^{(1)}
=
|x_{K_2}^{(2)}|
\prod_{p=0}^{K_2-1}
|\alpha_k^{(1)}-\alpha_p^{(2)}|.
\label{eq:GammaMR}
\end{equation}
Therefore, unlike conventional \ac{BMOCZ}, where the root stability depends only on intra-radius spacing, the proposed \ac{MR-BMOCZ} introduces an additional dependency on the geometric relationship between different radius groups. Substituting \eqref{eq:MRderivative} into \eqref{eq:rootPert} yields
\begin{equation}
\mathbb{E}\{|\Delta\alpha_k^{(1)}|^2\}
\approx
\frac{
N_0
\sum_{n=0}^{K}
|\alpha_k^{(1)}|^{2n}
}
{
|X_1'(\alpha_k^{(1)})|^2
|\Psi_k^{(1)}|^2
}.
\label{eq:MRrootPert}
\end{equation}
\begin{figure}[t!]
\centering
\resizebox{0.9\columnwidth}{!}{
\includegraphics{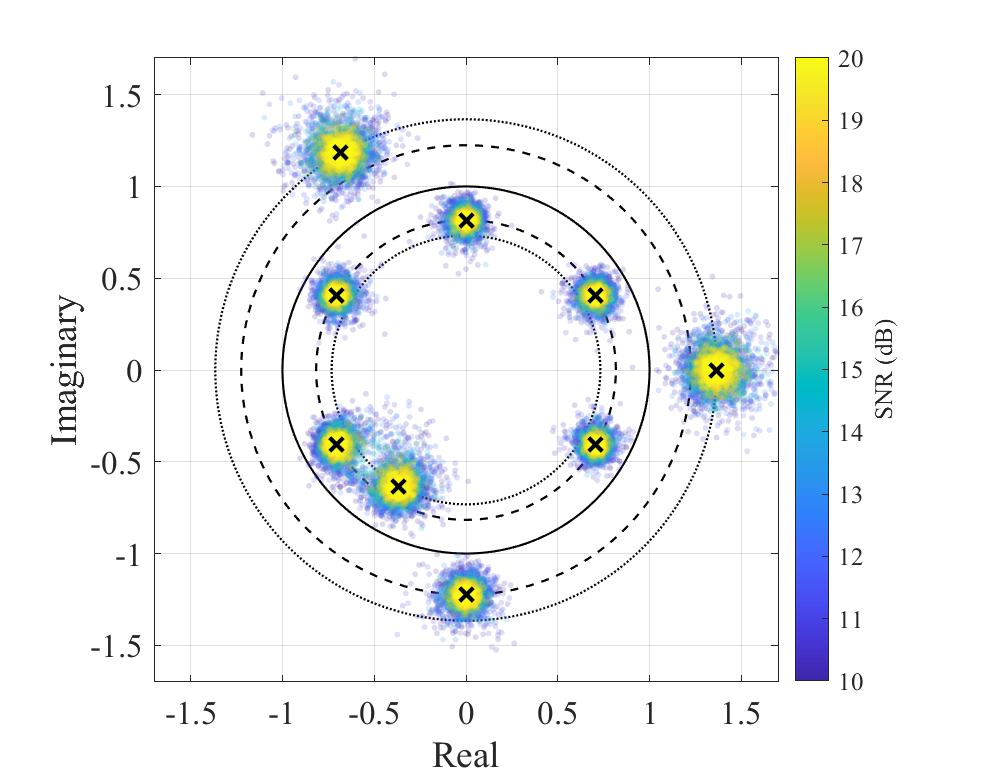}}
\caption{\ac{MR-BMOCZ} zeros perturbation in the complex plane for different \ac{SNR} values in an AWGN channel.}
\label{ZeroPert}
\end{figure}
From \eqref{eq:MRrootPert}, the perturbation of an \ac{MR-BMOCZ} zero is jointly determined by the conventional intra-radius spacing and the newly introduced inter-radius spacing. Consequently, the minimum separation between the two radius groups becomes an important parameter governing the detection performance. Let $d_{\min}^{(1,2)} = \min_{k,p} |\alpha_k^{(1)}-\alpha_p^{(2)}|$ denote the minimum inter-radius distance. Since every factor in \eqref{eq:GammaMR} is lower bounded by $d_{\min}^{(1,2)}$, it follows that
\begin{equation}
\Psi_k^{(1)}
\ge
|x_{K_2}^{(2)}|
\left(
d_{\min}^{(1,2)}
\right)^{K_2}.
\end{equation}
Substituting this bound into \eqref{eq:MRrootPert} gives
\begin{equation}
\mathbb{E}\{|\Delta\alpha_k^{(1)}|^2\}
\le
\frac{
N_0
\sum_{n=0}^{K}
|\alpha_k^{(1)}|^{2n}
}
{
|X_1'(\alpha_k^{(1)})|^2
|x_{K_2}^{(2)}|^2
\left(d_{\min}^{(1,2)}\right)^{2K_2}
}.
\label{eq:MRbound}
\end{equation}
The bound in \eqref{eq:MRbound} establishes a direct connection between the constellation geometry and the expected detection performance. In particular, increasing the minimum inter-radius separation increases the cross-radius robustness factor, reduces the expected root displacement, and improves the reliability of the \ac{DiZeT} decisions. Therefore, the angular optimization strategy developed in Lemma~1 not only improves the geometric separation of the constellation but also directly contributes to enhanced root stability. This behavior is illustrated in Fig.~\ref{ZeroPert}, where the estimated zero locations obtained from multiple noisy realizations form clusters around the transmitted zeros. Although the clusters become increasingly concentrated as the \ac{SNR} increases, the spread is not identical across all zeros. This unequal sensitivity is a direct consequence of the cross-radius robustness factor in \eqref{eq:GammaMR}, which varies according to the relative positioning of the radius groups.
\subsection{Complexity Analysis}
On the transmitter side, the proposed multi-radius scheme employs $I$ parallel streams to map the information-bearing zeros to the coefficients of the corresponding sub-polynomials. As shown in \cite{walk2020mocz}, the encoding complexity of conventional \ac{BMOCZ} is $\mathcal{O}(K^2)$. Considering \ac{MR-BMOCZ}, each polynomial construction in \eqref{multiplacationX} can be rewritten as
\begin{equation}
    X_i^{m} = (z-\alpha_m^{(i)})X_i^{m-1}(z).
\end{equation}
At recursion step $m$, we have $m$ multiplications and $m-1$ additions, hence the total number of multiplications $N_{mul}$ and additions $N_{add}$ required for all sub-polynomials is given as 
\begin{equation} \small
    N_{mul}= \sum_{m=1}^{K_i}m = \frac{K_i(K_i+1)}{2} \;,\; N_{add}= \sum_{m=1}^{K_i}(m-1) = \frac{K_i(K_i-1)}{2}.
\end{equation}
Thus, each sub-polynomial requires a complexity of $C_i = \mathcal{O}(K_i^2)$, and as all $I$ groups are constructed independently, the resulting complexity is given as $\mathcal{O}(\sum_{i=1}^{I} K_i^2)$. After the sub-polynomial encoding stage, the resulting time-domain signal is given in \eqref{eqautocorr} as the convolution of the resulting sub-polynomial time-domain signals. By setting $S_i = \sum_{j=1}^i K_i$, combining the first $i-1$ polynomials with the $i^{th}$ polynomial requires a convolution of degree $S_{i-1}$ and $K_i$, hence the complexity of this operation is given as $\mathcal{O}(S_{i-1}K_i)$. Thus, the total cost of convolving all sub-polynomials will be given as $\mathcal{O}(\sum_{i=2}^IS_{i-1}K_i)$. Using $K=\sum_iK_i$ we get
\begin{equation}
\sum_{i=2}^{I} S_{i-1}K_i = \sum_{j<i} K_jK_i = \frac{1}{2}\left[K^2-\sum_{i=1}^{I}K_i^2\right].
\end{equation}
The resulting total time complexity of the \ac{MR-BMOCZ} scheme is then given as
\begin{equation}
\begin{aligned}
C_{\mathrm{enc,total}}
&=
\mathcal{O}\left(\sum_{i=1}^{I}K_i^2 + \frac{K^2-\sum_{i=1}^{I}K_i^2}{2}\right) \\
&=
\mathcal{O}(K^2).
\end{aligned}
\end{equation}
With actual parallel hardware, the latency is given as $T = \mathcal{O}(\max_i K_i^2)$ compared to conventional latency given as $T = \mathcal{O}(K^2)$.

On the receiver side, the received polynomial has degree $N = K+L-1$ where $L$ is the channel length. The \ac{DiZeT} decoder evaluates the received polynomial at all possible zero locations, hence the resulting decoding complexity for conventional \ac{BMOCZ} is $\mathcal{O}(K^2)$ \cite{walk2020mocz}. Considering \ac{MR-BMOCZ}, for each radius group, the \ac{DiZeT} tests $2K_i$ candidate zero locations, hence the $i$-th group requires $C_i = \mathcal{O}(2K_iN)$. Summing across all groups yields $C = \mathcal{O}(2N\sum_{i=1}^IK_i)$ which is the same as $\mathcal{O}(2NK)$, which results in a time complexity of $C_{total} = \mathcal{O}(K(K+L))$. For $L<<K$, we get a time complexity of $\mathcal{O}(K^2)$, which is equivalent to the conventional time complexity. With actual parallel hardware, the latency is given as $T = \mathcal{O}((K+L)\max_iK_i)$ compared to conventional latency given as $T = \mathcal{O}(K(K+L))$.

\section{Simulation and Numerical Results} \label{simulationsection}
In this section, we evaluate the proposed \ac{MR-BMOCZ} scheme through numerical simulations. We first establish its baseline \ac{BER} performance relative to conventional \ac{BMOCZ}, and subsequently compare its robustness against \ac{J-BMOCZ} and \ac{S-BMOCZ} in the presence of timing offsets. Unless stated otherwise, \ac{MR-BMOCZ} employs $K_1=7$ and $K_2=14$, and all waveforms are decoded using the \ac{DiZeT} receiver. Transmission is simulated over both \ac{AWGN} and fading channels. For the \ac{AWGN} case, each sample of the noise sequence is independently drawn from $\mathcal{CN}(0,N_0)$. In all error rate simulations, the binary message vector is selected uniformly from the set of all possible messages. To assess synchronization robustness, the \ac{TO} is varied across the considered observation interval for each transmitted packet, and to simulate the impact of \ac{CFO} impairment, we sample $\varepsilon$ from the uniform distribution $\mathcal{U}_{[0,\frac{\pi}{7})}$ to be consistent with the assumption made in Section \ref{sec:sync}.
\subsection{TO performance}
\begin{figure}[!t]
    \centering
    \begin{subfigure}[b]{0.44\textwidth}
        \centering
        \includegraphics[width=\linewidth]{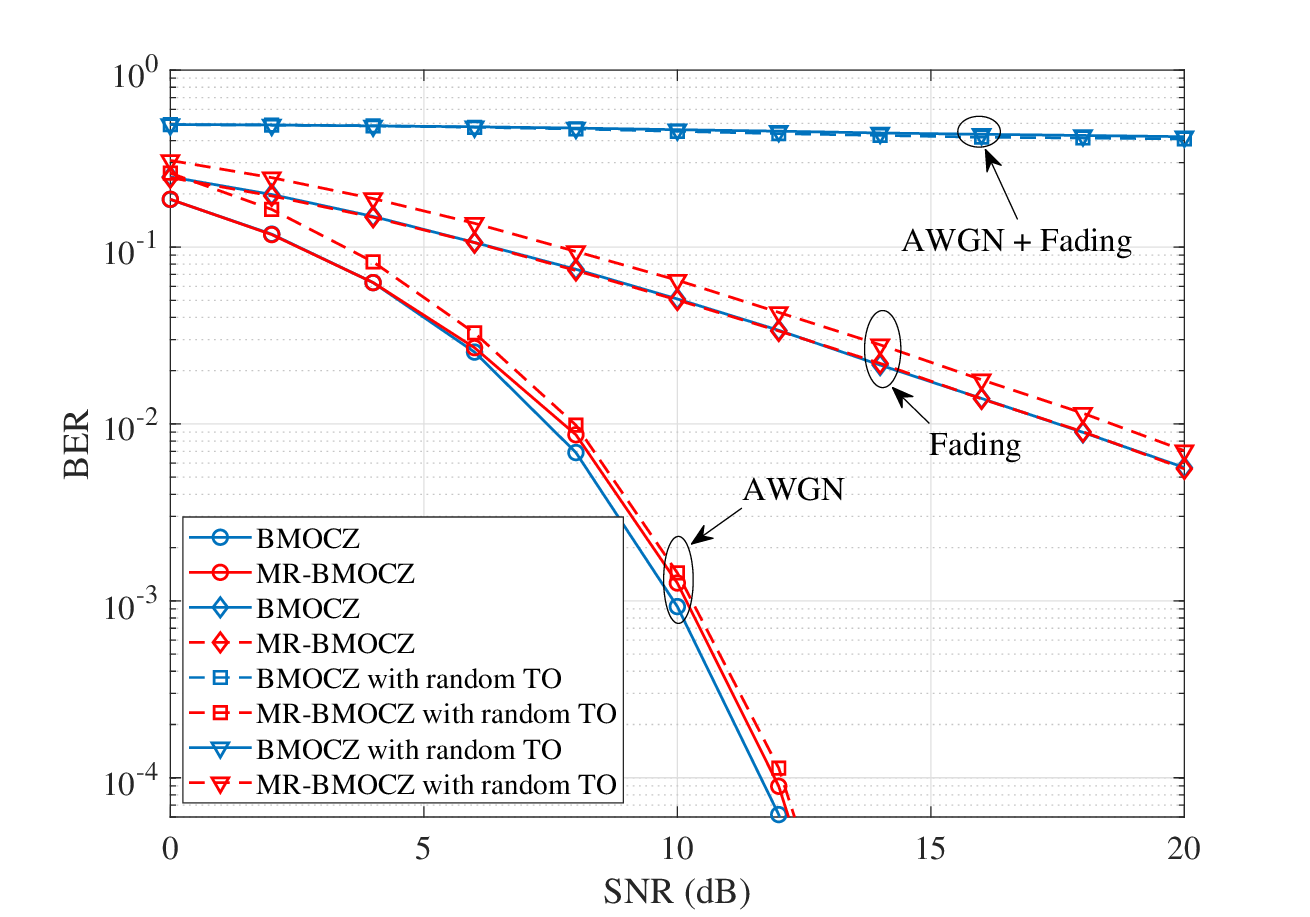}
        \caption{\ac{BER} curves.}
        \label{fig:BER}
    \end{subfigure}
    \vfill
    \begin{subfigure}[b]{0.44\textwidth}
        \centering
        \includegraphics[width=\linewidth]{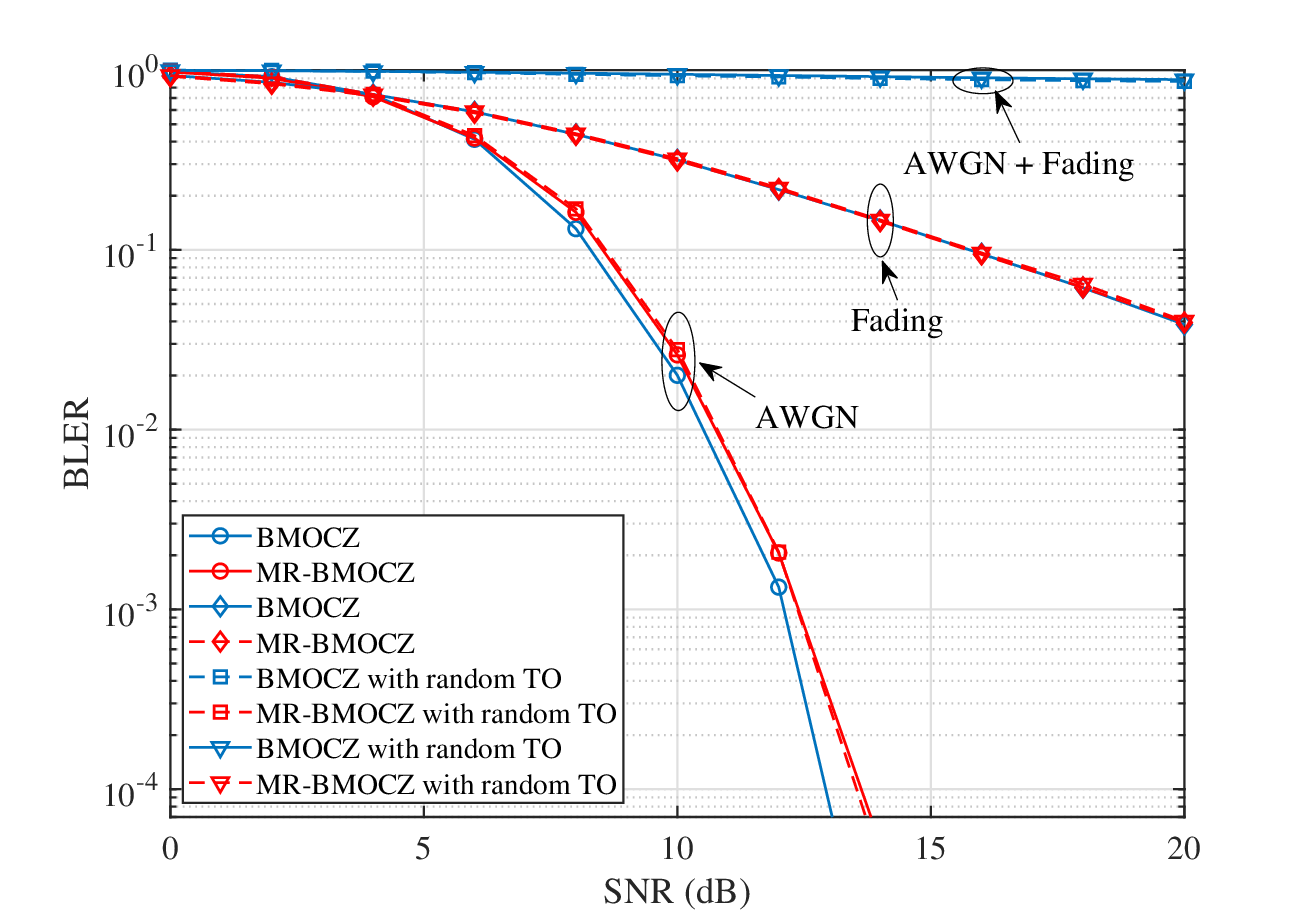}
        \caption{BLER curves.}
        \label{fig:BLER}
    \end{subfigure}
    \caption{Performance of conventional \ac{BMOCZ} and the proposed \ac{MR-BMOCZ}.}
    \label{fig:errorPerformance}
\end{figure}

Fig.~\ref{fig:BER} reports the simulated \ac{BER} in \ac{AWGN} and flat-fading channels for \ac{MR-BMOCZ} and \ac{BMOCZ}. Under perfect synchronization, both schemes achieve comparable \ac{BER} performance in \ac{AWGN} and fading channels. In the \ac{AWGN} case, the two schemes achieve a rapid decrease in \ac{BER} as the \ac{SNR} increases, reaching error rates below $10^{-4}$ at high \ac{SNR} values. A similar trend is observed in the fading channel, with moderate performance degradation for the multi-radius scheme relative to the conventional one, caused by multipath propagation. These results show that the multi-radius construction retains the detection performance of conventional \ac{BMOCZ} despite the modified zero structure. The impact of timing synchronization errors is also illustrated by considering a random \ac{TO} and the corresponding \ac{TO} correction procedure.  Conventional \ac{BMOCZ} experiences a significant degradation in \ac{BER} both in \ac{AWGN} and fading channels where an elevated error floor is observed, whereas the performance of \ac{MR-BMOCZ} approaches that of the perfectly synchronized case. This behavior is a direct consequence of the richer autocorrelation structure of \ac{MR-BMOCZ}, which provides additional timing references for the synchronization algorithm.  As a result, \ac{MR-BMOCZ} is considerably less sensitive to timing errors while maintaining a \ac{BER} performance close to that of conventional \ac{BMOCZ}. The  \ac{BLER} results in Fig.~\ref{fig:BLER} follow the same overall behavior where under perfect synchronization, \ac{MR-BMOCZ} and conventional \ac{BMOCZ} achieve nearly identical block-level performance in both \ac{AWGN} and flat-fading channels. However, a random \ac{TO} causes a severe \ac{BLER} degradation for the conventional case since an inaccurate timing estimate affects the entire received zero pattern. After applying the \ac{TO} correction technique, \ac{MR-BMOCZ} achieves a lower \ac{BLER} in the fading case and remains considerably closer to the perfectly synchronized benchmark. As a result, the additional autocorrelation peaks offered by the multi-radius structure improve packet-level robustness against timing uncertainty.
\subsection{Normalized CFO performance}
\begin{figure}[t!]
\centering
\resizebox{0.9\columnwidth}{!}{
\includegraphics{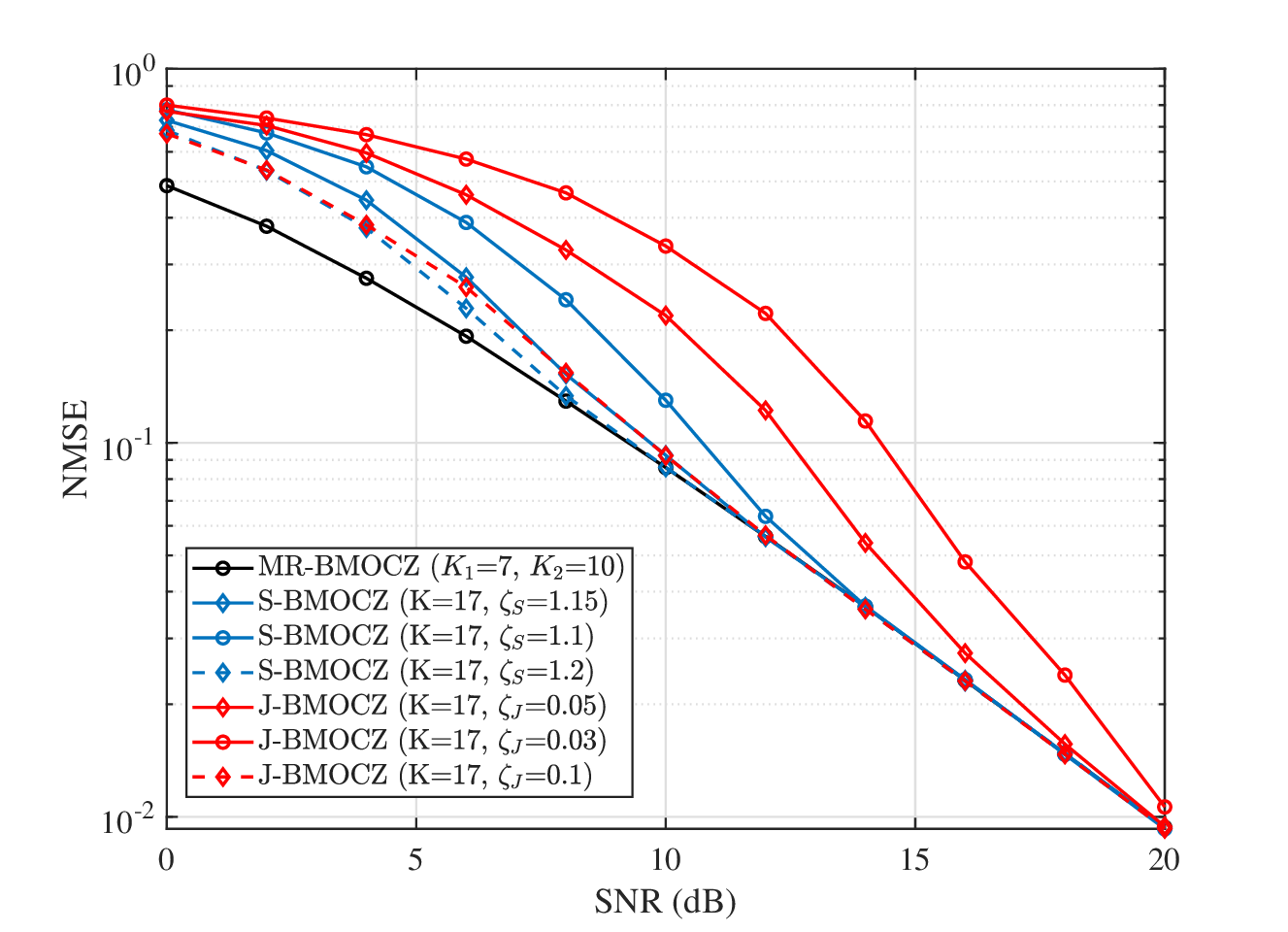}}
\caption{\ac{NMSE} Performance of \ac{CFO} Estimation versus SNR.}
\label{RMSE}
\end{figure}

Fig.~\ref{RMSE} compares the \ac{NMSE} of the \ac{CFO} estimate for the proposed \ac{MR-BMOCZ} scheme with those of the benchmark \ac{S-BMOCZ} \cite{perre2025smooshed} and \ac{J-BMOCZ} \cite{huggins2025fourier} approaches. Both benchmark schemes introduce an asymmetry factor, denoted by $\zeta$, to produce a unique spectral signature that can be exploited  for \ac{CFO} detection and correction purposes. Detailed derivations are provide in \cite{perre2025smooshed} and \cite{huggins2025fourier}. Across the entire \ac{SNR} range, the proposed \ac{MR-BMOCZ} achieves the lowest \ac{NMSE} among the considered schemes. Although increasing the asymmetry parameter $\zeta$ improves the performance of both \ac{S-BMOCZ} and \ac{J-BMOCZ}, the proposed scheme maintains a clear advantage due to the presence of multiple spectral components associated with the different radius groups. However, note that increasing the asymmetry parameter $\zeta$ reduces the \ac{BER} performance of these two aforementioned schemes, whereas our proposed scheme maintains a fixed \ac{BER} versus \ac{SNR} performance. As the \ac{SNR} increases, all schemes converge to a common performance determined by the search-grid resolution. Nevertheless, \ac{MR-BMOCZ} exhibits a faster \ac{NMSE} decay in the low and moderate \ac{SNR} regions, where \ac{CFO} estimation accuracy is most critical.
\begin{figure}[t!]
\centering
\resizebox{0.9\columnwidth}{!}{
\includegraphics{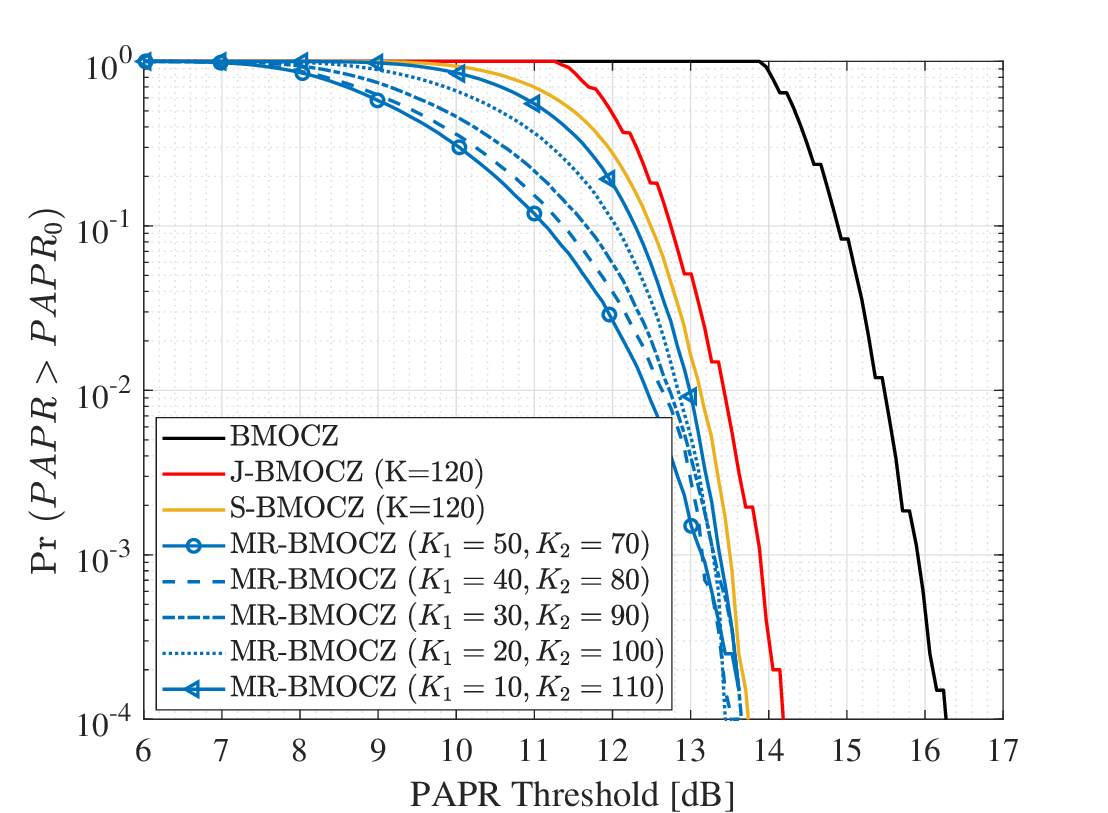}}
\caption{CCDF of the PAPR for BMOCZ, J-BMOCZ, S-BMOCZ and the proposed MR-BMOCZ schemes.}
\label{PAPRfig}
\end{figure}
\begin{figure}[t]
\centering
\resizebox{0.9\columnwidth}{!}{
\includegraphics{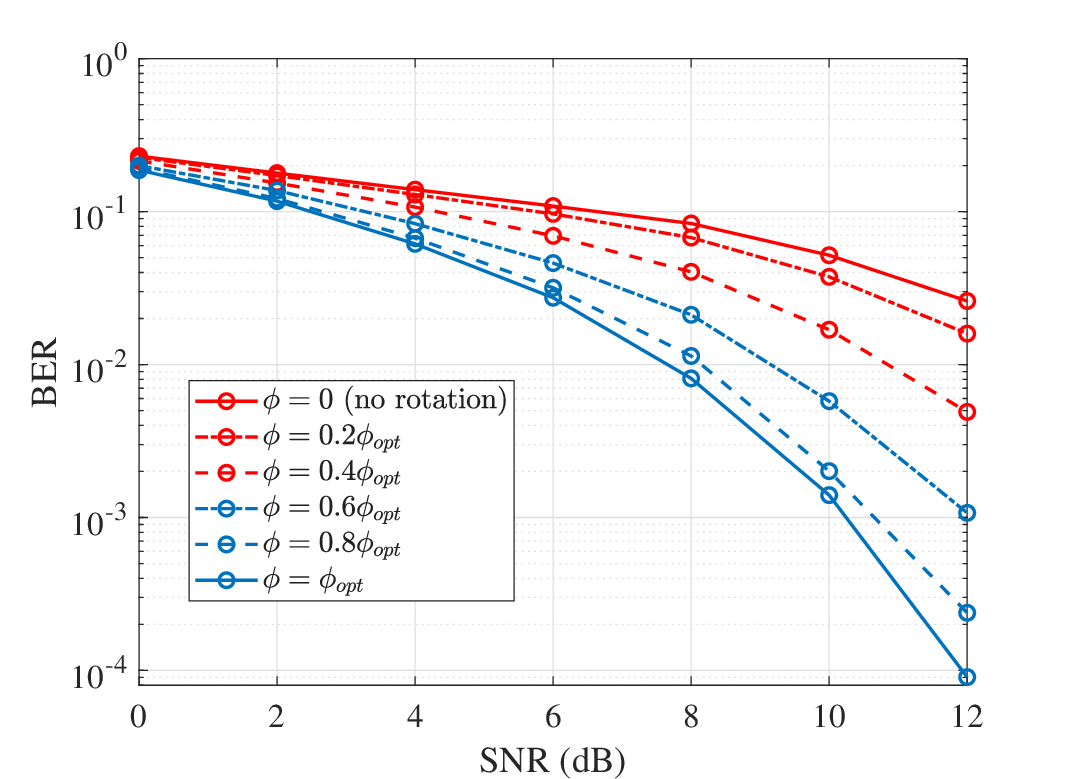}}
\caption{BER comparison for various phase rotation angles.}
\label{PhaseBER}
\end{figure}
\subsection{PAPR performance}
\begin{figure}[t]
\centering
\resizebox{0.9\columnwidth}{!}{
\includegraphics{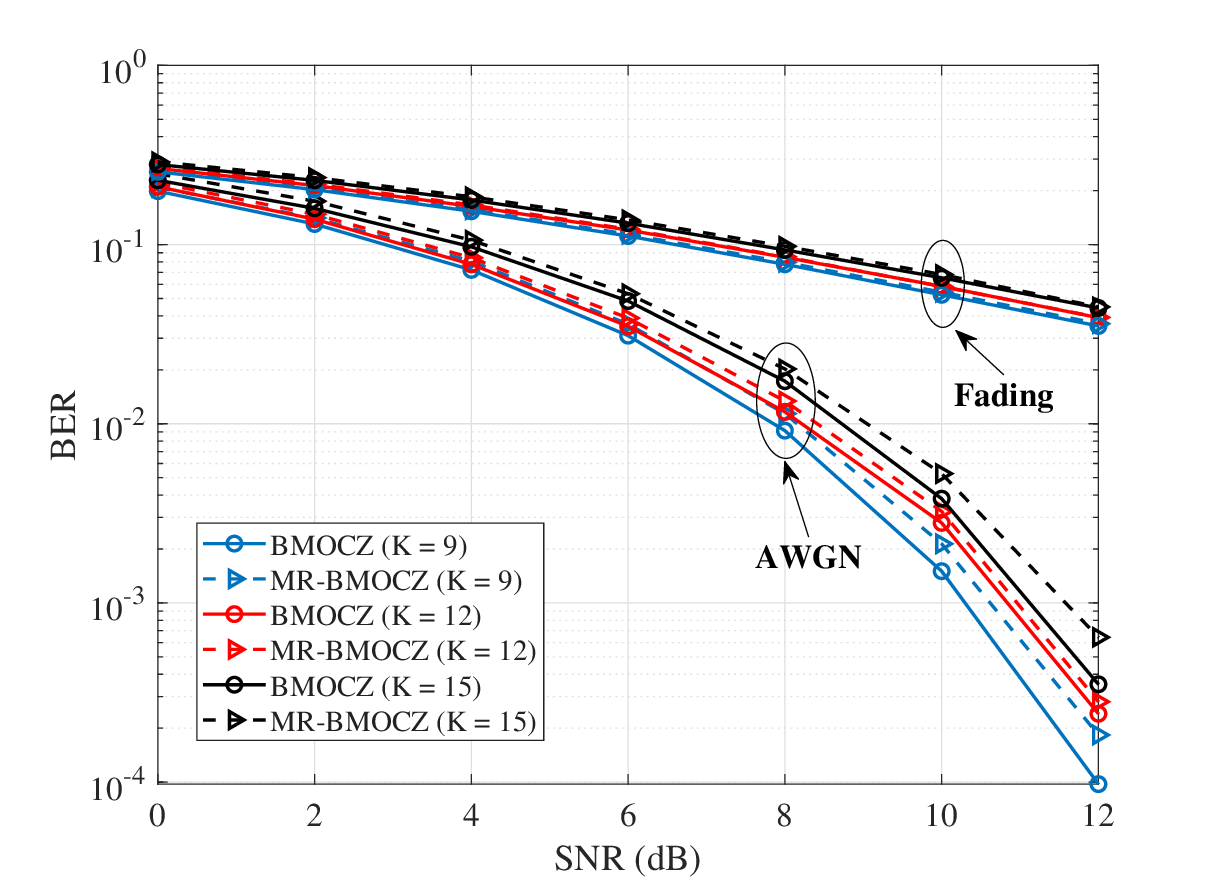}}
\caption{\ac{BER} performance comparison for different values of $K$ in \ac{AWGN} and fading channel.}
\label{diffKBER1}
\end{figure}
The \ac{PAPR} \ac{CCDF} results are presented in Fig.~\ref{PAPRfig}. The proposed \ac{MR-BMOCZ} consistently shifts the \ac{CCDF} curves toward lower \ac{PAPR} values, corresponding to a lower probability of large-amplitude waveform peaks. This behavior follows directly from the multi-radius construction, where the transmitted sequence is generated through the convolution of several lower-order Huffman sequences rather than a single high-order polynomial. As a result, the signal energy is distributed more uniformly, mitigating energy concentration in a small number of samples. Furthermore, the results show that the \ac{PAPR} reduction becomes more pronounced as the radius partition becomes increasingly balanced. In particular, configurations with comparable values of $K_1$ and $K_2$ achieve the lowest \ac{PAPR}, whereas highly unbalanced partitions gradually approach the behavior of conventional \ac{BMOCZ}. It can also be observed that the proposed \ac{MR-BMOCZ} outperforms not only conventional \ac{BMOCZ} but also the recently proposed \ac{J-BMOCZ}~\cite{huggins2025fourier} and \ac{S-BMOCZ}~\cite{perre2025smooshed}, demonstrating that multi-radius signaling provides an effective mechanism for reducing \ac{PAPR} while simultaneously preserving the synchronization advantages discussed previously.
\subsection{BER performance}
The impact of the phase rotation on the BER performance is illustrated in Fig.~\ref{PhaseBER}. It can be observed that the phase rotation given by \eqref{eq:optimalphase} achieves the lowest BER among the tested configurations. This result confirms the strong correlation between constellation geometry and detection reliability, since larger angular distances reduce the probability of incorrect zero classification under noise.
Fig.~\ref{diffKBER1} displays a comparison of \ac{BER} performance for \ac{MR-BMOCZ} and conventional \ac{BMOCZ} in \ac{AWGN} and flat-fading environments for various values of $K$. For the \ac{AWGN} channel, the \ac{BER} curves remain relatively unchanged across the considered packet lengths, indicating that the proposed construction is insensitive to the choice of $K$. As $K$ takes larger values, a relatively minor performance gap is noticed between \ac{MR-BMOCZ} and conventional \ac{BMOCZ}, suggesting that the introduction of additional radius groups does not significantly alter the underlying detection behavior.
\begin{figure}[t]
\centering
\resizebox{0.9\columnwidth}{!}{
\includegraphics{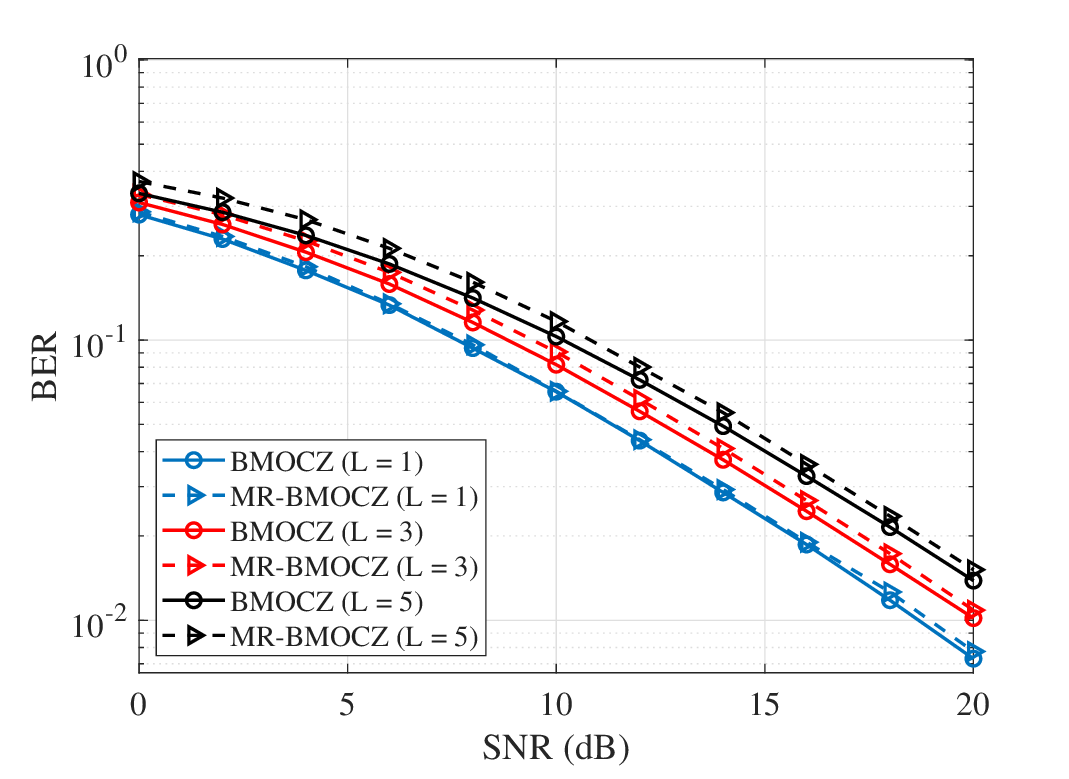}}
\caption{\ac{BER} performance comparison between using \ac{DiZeT} detection for $K=9$ for various channel lengths $L$.}
\label{diffLBER}
\end{figure}
In the flat-fading case, the \ac{BER} performance remains nearly unchanged across the considered values of $K$. The proposed scheme follows the same performance trend as the conventional \ac{BMOCZ} scheme, suggesting that the multi-radius construction does not introduce a significant performance penalty. This suggests that, in the flat-fading scenario, the relative difference between the two zero placement strategies is minimized.
In Fig.~\ref{diffLBER}, in the frequency-selective fading scenario, the performance of both schemes is evaluated by varying the channel length $L$ while maintaining a fixed message length. In this scenario, both schemes exhibit a similar degradation as the channel length increases. As expected, the \ac{BER} performance of both schemes degrades as the channel length is increased, since longer channel responses introduce additional multipath-induced distortions, which increase the difficulty of blind detection.
\section{Conclusion and Future Work} \label{conc}
This paper introduced \ac{MR-BMOCZ}, a synchronization-oriented extension of conventional \ac{BMOCZ} using a structured multi-radius zero constellation. By distributing information-bearing zeros across multiple conjugate-reciprocal radii, the proposed scheme produces additional deterministic autocorrelation peaks that are exploited for blind synchronization. The resulting waveform structure also promotes balanced energy distribution, leading to a reduced \ac{PAPR} compared with \ac{BMOCZ}-based schemes. Special techniques to estimate \ac{TO} and \ac{CFO} were developed to leverage the unique autocorrelation and spectral signatures of the multi-radius structure. As a result, the receiver preserves the non-coherent nature of conventional \ac{BMOCZ} while improving time synchronization performance and enhancing frequency-domain signature. Simulation results in \ac{AWGN} and fading channels have shown notable gains in \ac{TO} estimation reliability and \ac{CFO} estimation accuracy, yielding a lower \ac{NMSE} than conventional \ac{BMOCZ}-based synchronization approaches while maintaining a comparable \ac{BER} and \ac{BLER} performance. Future work will investigate the joint optimization of radius allocation, message partitioning and zero placement to improve detection.

\section*{Appendix A}
\section*{\centering Proof of Lemma 1}
Consider two angular grids defined as
\begin{equation}
\Theta_i =
\left\{
\theta_{k}^{(i)}
=
\frac{2\pi k}{K_i}
+
\Phi_i^{(0)}
\;:\;
k=0,\ldots,K_i-1
\right\},
\end{equation}
and
\begin{equation}
\Theta_j =
\left\{
\theta_{\ell}^{(j)}
=
\frac{2\pi \ell}{K_j}
+
\Phi_j^{(0)}
\;:\;
\ell=0,\ldots,K_j-1
\right\}.
\end{equation}
The relative phase offset between the two grids is defined as
\begin{equation}
\Delta\Phi_{i,j}
\triangleq
\left[
\Phi_j^{(0)}
-
\Phi_i^{(0)}
\right]_{2\pi}.
\end{equation}
An angular collision occurs if there exist integers $k$ and $\ell$ such that $\left[
\theta_{k}^{(i)}
\right]_{2\pi}
=
\left[
\theta_{\ell}^{(j)}
\right]_{2\pi}$. Hence
\begin{equation}
\Delta\Phi_{i,j}
=
\left[
2\pi
\left(
\frac{k}{K_i}
-
\frac{\ell}{K_j}
\right)
\right]_{2\pi}.
\end{equation}
Let
\begin{equation}
G_{i,j}
=
\operatorname{lcm}(K_i,K_j)
=
\frac{K_iK_j}{\operatorname{gcd}(K_i,K_j)},
\end{equation}
where $\operatorname{gcd}(\cdot,\cdot)$ denotes the greatest common divisor. Since both fractions $k/K_i$ and $\ell/K_j$ can be expressed using the common denominator $G_{i,j}$, their difference can be written as
\begin{equation}
\frac{k}{K_i}
-
\frac{\ell}{K_j}
=
\frac{g}{G_{i,j}},
\end{equation}
for some integer $g\in\{0,1,\ldots,G_{i,j}-1\}$. Consequently, the set of relative phase offsets producing angular collisions is
\begin{equation}
\mathcal{C}_{i,j}
=
\left\{
2\pi\frac{g}{G_{i,j}}
\;:\;
g=0,1,\ldots,G_{i,j}-1
\right\}.
\end{equation}
The elements of $\mathcal{C}_{i,j}$ partition the unit circle into $G_{i,j}$ equally spaced points. The minimum angular distance to this set is maximized when the relative phase offset is located in-between two consecutive elements of $\mathcal{C}_{i,j}$. Therefore, the optimal relative phase offset is
\begin{equation}
\Delta\Phi_{i,j}^{\star}
=
\frac{1}{2}
\left(
\frac{2\pi}{G_{i,j}}
\right)
=
\frac{\pi}{G_{i,j}}.
\end{equation}
This choice maximizes the worst-case angular separation between the points in $\Theta_i$ and $\Theta_j$, completing the proof.
\section*{Appendix B}
\section*{\centering Proof of Lemma 2}

Consider the $i$-th angular grid
\begin{equation}
\Theta_i
=
\left\{
\theta_k^{(i)}
=
\frac{2\pi k}{K_i}
+
\Phi_i^{(0)}
\;:\;
k=0,\ldots,K_i-1
\right\}.
\end{equation}

A CFO-induced phase rotation $\varepsilon$ maps each angular location according to $\theta \mapsto \left[\theta-\varepsilon\right]_{2\pi}$. Accordingly, the rotated $i$th grid is
\begin{equation}
\Theta_i-\varepsilon
=
\left\{
\left[
\theta_k^{(i)}-\varepsilon
\right]_{2\pi}
\;:\;
k=0,\ldots,K_i-1
\right\}.
\end{equation}

Since a CFO-induced rotation preserves the radius of each zero, the overall multi-radius constellation remains invariant only if every radius group remains invariant individually. Thus, $\Theta_i$ is invariant if and only if there exists an integer $m_i$ such that $\left[
\theta_k^{(i)}-\varepsilon
\right]_{2\pi}
=
\left[
\theta_{k-m_i}^{(i)}
\right]_{2\pi},
\qquad
\forall k$. Thus
\begin{equation}
\left[
\frac{2\pi k}{K_i}
+
\Phi_i^{(0)}
-
\varepsilon
\right]_{2\pi}
=
\left[
\frac{2\pi(k-m_i)}{K_i}
+
\Phi_i^{(0)}
\right]_{2\pi}.
\end{equation}
Therefore,
\begin{equation}
\varepsilon
\equiv
\frac{2\pi m_i}{K_i}
\pmod{2\pi}.
\end{equation}
Hence, the $i$-th angular grid is invariant if and only if
$\varepsilon
\in
\left\{
\frac{2\pi m_i}{K_i}
\;:\;
m_i\in\mathbb{Z}
\right\}$.
For the union constellation $\Theta_{\mathrm{MR}}
=
\bigcup_{i=1}^{I}\Theta_i$
to remain invariant, the above condition must hold simultaneously for all $i=1,\ldots,I$. Therefore,
\begin{equation}
\varepsilon
\in
\bigcap_{i=1}^{I}
\left\{
\frac{2\pi m_i}{K_i}
\;:\;
m_i\in\mathbb{Z}
\right\}.
\end{equation}
The intersection of these angular lattices is given by
\begin{equation}
\varepsilon
=
\frac{2\pi q}
{\operatorname{gcd}(K_1,\ldots,K_I)},
\qquad
q\in\mathbb{Z},
\end{equation}
which establishes the set of CFO-induced phase rotations that leave the multi-radius zero constellation invariant. The minimum nonzero invariant rotation is obtained for $q=1$, completing the proof.

\bibliographystyle{IEEEtran}
\bibliography{mybib}
\vfill
\end{document}